\documentclass[pra,unsortedaddress,12pt]{revtex4-1}
\usepackage[english]{babel}
\usepackage[utf8]{inputenc}
\usepackage{graphicx}
\usepackage{comment}
\usepackage{hyperref}
\usepackage{color}
\usepackage{amsmath,amssymb,amsfonts} 
\usepackage{subfig}

\input{Qcircuit} 

\begin{document}

\title{Quantum computing approach to non-relativistic and relativistic molecular energy calculations}

\author{Libor Veis}
\email{libor.veis@jh-inst.cas.cz}
\affiliation{\mbox{J. Heyrovsk\'{y} Institute of Physical Chemistry, Academy of Sciences of the Czech Republic, v.v.i.}, Dolej\v{s}kova 3, 18223 Prague 8, Czech Republic \\}
\affiliation{\mbox{Charles University in Prague, Faculty of Science, Department of Physical and Macromolecular Chemistry}, Hlavova 8, 12840 Prague 2, Czech Republic}

\author{Ji\v{r}\'{i} Pittner}
\email{jiri.pittner@jh-inst.cas.cz, corresponding author}
\affiliation{\mbox{J. Heyrovsk\'{y} Institute of Physical Chemistry, Academy of Sciences of the Czech Republic, v.v.i.}, Dolej\v{s}kova 3, 18223 Prague 8, Czech Republic}

\begin{abstract}
\tableofcontents
\end{abstract}
\maketitle

\section{Overview}
An \textit{exact} simulation of quantum systems on a classical computer is computationally hard. The problem lies in the dimensionality of the Hilbert space needed for the description of our system that in fact grows exponentially with the size of this system. No matter if we simulate the dynamics or calculate some static property e.g. the energy, this limitation is always present. Richard Feynman came up with an alternative to the classical simulation \cite{feynman_1982}. His idea was to convert the aforementioned drawback of quantum systems into their benefit. He suggested to map the Hilbert space of a studied quantum system on another one (both of them being exponentially large) and thus to \textit{efficiently} simulate a quantum system on another one (i.e. on a quantum computer).

This was the original idea of quantum computers. There is no doubt that quantum computing is nowadays a well-established  discipline of computer science. Apart from the \textit{efficient} simulation of quantum systems \cite{lloyd_1996, zalka_1998, ortiz_2001, somma_2002, abrams_1997}, other interesting problems where quantum computers could beat their classical counterparts have been discovered. The most famous examples are integer factorization for which quantum computers supply an exponential speedup \cite{shor_1994, shor_1997} or database search \cite{grover_1997}. However, for our purposes the \textit{efficient} (polynomially scaling) quantum algorithm of Abrams and Lloyd for obtaining eigenvalues of local Hamiltonians \cite{abrams_1999} is particularly important. 

The first paper connecting quantum computation and chemistry was published by Lidar and Wang \cite{lidar_1999} and concerned the \textit{efficient} calculations of thermal rate constants of chemical reactions. This work in fact founded the new field of computational chemistry that is the subject of this book, namely the ``computational chemistry on quantum computers". Aspuru-Guzik et al. in their seminal article \cite{aspuru-guzik_2005} reduced the number of quantum bits (qubits) needed by the Abrams and Lloyd's algorithm \cite{abrams_1999} and applied it to molecular ground state energy calculations. Since these two pioneering works, other papers involving energy calculations of excited states \cite{wang_2008}, quantum chemical dynamics \cite{kassal_2008}, calculations of molecular properties and geometry optimizations \cite{kassal_2009}, state preparations \cite{ward_2009,wang_2009} or global minima search \cite{zhu_2009} were published. The list of all chemical applications for quantum computers is quite rich and an interested reader can find a comprehensive review in \cite{kassal_review}.

Aspuru-Guzik et al. \cite{aspuru-guzik_2005} also proposed that quantum computers with tens of (noise free) qubits would already exceed the limits of classical full configuration interaction (FCI) calculations. This is in contrast to other quantum algorithms, e.g. the Shor's algorithm \cite{shor_1994, shor_1997} for integer factorization would for practical tasks in cryptography require thousands of qubits. For this reason, calculations and simulations of quantum systems will belong to the first practical applications of quantum computers. Recent proof-of-principle few-qubit experiments covering energy calculations of the hydrogen molecule \cite{lanyon_2010, du_2010} or Heisenberg spin model \cite{li_2011} and the simulation of a chemical reaction dynamics \cite{lu_2011} confirm that interesting applications might be just behind the door.

The aim of this article is to review quantum algorithms for exact molecular energy calculations (within a finite one-particle basis set), i.e. quantum analogues of the FCI calculations. On a classical computer, a computational cost of the FCI method scales exponentially with the size of the system. This fact stems from the dimension of the Hilbert space in which we diagonalize the Hamiltonian matrix and it is the reason why this method is limited only to the smallest systems (diatomics, triatomics). For example, in the non-relativistic case, the number of Slater determinants that build up the FCI wave function for a closed-shell system with $n$ electrons in $m$ orbitals is equal to

\begin{equation}
  N_{\rm{non-rel.}} = \left( \begin{array}{c}  m \\ n/2 \end{array} \right)^2.
  \label{nrdets}
\end{equation}

\noindent
It is evident that this number grows into huge values with increasing $m$ very quickly. On a quantum computer on the other hand, it has been shown \cite{lanyon_2010, whitfield_2010} and will be discussed later in the article that the FCI cost has a polynomial scaling [$\mathcal{O}(m^5)$], therefore it is \textit{exponentially} faster.

The structure of this article is as follows: First we very briefly give the necessary basics of quantum computing including the quantum Fourier transform (QFT), the phase estimation algorithm (PEA) and their semiclassical variants (Section \ref{Quantum computing background}). Despite being textbook topics \cite{nielsen_chuang}, we believe that it may be convenient for readers from the quantum chemistry community to make this article more self-contained. Section \ref{Quantum full configuration interaction method} presents details of the quantum algorithm for the FCI method (qFCI). In Section \ref{NR}, we show its applications to non-relativistic computations, namely the computations of the ground and excited state energies of the methylene molecule (CH$_{2}$). We then generalize this approach to the relativistic 4-component (4c) calculations and apply it to the problem of spin-orbit splitting in the SbH molecule in Section \ref{R}.

\section{Quantum computing background}
\label{Quantum computing background}

In this section we briefly mention the necessary quantum computing background. For a more detailed description, we refer the reader e.g. to the excellent book by Nielsen and Chuang \cite{nielsen_chuang}.

\subsection{Quantum Fourier transform}
The classical discrete Fourier transform takes as an input a vector of complex numbers ($x_0,\ldots,x_{N-1}$) and outputs the elements of another vector ($y_0,\ldots,y_{N-1}$) according to the equation

\begin{equation}
  y_k = \frac{1}{\sqrt{N}} \sum_{j=0}^{N-1} x_j e^{2 \pi ijk/N}.
  \label{dft}
\end{equation}

\noindent
Similarly, the quantum Fourier transform (QFT) operates on an orthonormal basis of $n$ qubits: $\ket{0} \ldots \ket{2^n-1}$ and is defined as an operator $\hat{U}_{\rm{QFT}}$ 

\begin{equation}
  \hat{U}_{\rm{QFT}} \ket{k} = \frac{1}{\sqrt{N}} \sum_{j=0}^{N-1} e^{2 \pi ijk/N} \ket{j}, \qquad N = 2^n,
  \label{qft_eq}
\end{equation}

\noindent
where the kets are numbered by a binary representation of integers

\begin{equation}
  \begin{array}{cc}
    \ket{j} = \ket{j_{n} \dots j_{1}}, \qquad j_i \in \{ 0,1 \},  \\[0.2cm]
    \displaystyle{j = \sum_{i=1}^{n} j_{i} \cdot 2^{i-1} .}
  \end{array}
\end{equation}

\begin{figure}[!ht] 
  \mbox{
  \hskip -0.2cm 
  \Qcircuit @C=0.25em @R=0.5em { 
    \lstick{\ket{q_{n}}} & \gate{H} & \ctrl{1} & \ctrl{2} & \qw & & \cdots & & & \ctrl{4} & \ctrl{5} & \qw & \qw & \qw & \qw & \qw & \qw & \qw &\qw & \qw & \qw & \qw & \qw & \qw & \qw & \qw & \qw & \qw & \qw & \qw & \qw & \qw & \qw & \qw & \qw & \qw & \qw & \qw & \qw & \rstick{\ket{\tilde{q}_1}} \\
    \lstick{\ket{q_{n-1}}} & \qw & \gate{R_2} & \qw & \qw & & \cdots & & & \qw & \qw & \qw & \gate{H} & \ctrl{1} & \qw & & \cdots & & & \ctrl{3} & \ctrl{4} & \qw & \qw & \qw & \qw & \qw & \qw & \qw & \qw & \qw & \qw & \qw & \qw & \qw & \qw & \qw & \qw & \qw & \qw & \rstick{\ket{\tilde{q}_2}}\\
    \lstick{\ket{q_{n-2}}} & \qw & \qw & \gate{R_3} & \qw & & \cdots & & & \qw & \qw & \qw & \qw & \gate{R_{2}} & \qw & & \cdots & & & \qw & \qw & \gate{H} & \qw & & \cdots & & & \ctrl{2} & \ctrl{3} & \qw & \qw & \qw & \qw & \qw & \qw & \qw & \qw & \qw & \qw & \rstick{\ket{\tilde{q}_3}} \\
    & \vdots & & & & & & & & & & & \vdots & & & & & & & & & \vdots & & & & & & & & & & & & & & & & \vdots \\
    \lstick{\ket{q_{2}}} & \qw & \qw & \qw & \qw & \qw & \qw & \qw & \qw & \gate{R_{n-1}} & \qw & \qw & \qw & \qw & \qw & \qw & \qw & \qw & \qw & \gate{R_{n-2}} & \qw & \qw & \qw & \qw & \qw & \qw & \qw & \gate{R_{n-3}} & \qw & \qw & & \cdots & & & \qw & \gate{H} & \ctrl{1} & \qw & \qw &  \rstick{\ket{\tilde{q}_{n-1}}} \\
    \lstick{\ket{q_{1}}} & \qw & \qw & \qw & \qw & \qw & \qw & \qw & \qw & \qw & \gate{R_{n}} & \qw & \qw & \qw & \qw & \qw & \qw & \qw & \qw & \qw & \gate{R_{n-1}} & \qw & \qw & \qw & \qw & \qw & \qw & \qw & \gate{R_{n-2}} & \qw & & \cdots & & & \qw & \qw & \gate{R_2} & \gate{H} & \qw & \rstick{\ket{\tilde{q}_{n}}}
  }}
  \caption{The quantum Fourier transform circuit. Note that qubits of the result are in a reversed order after the application of this circuit.}
  \label{qft}
\end{figure}
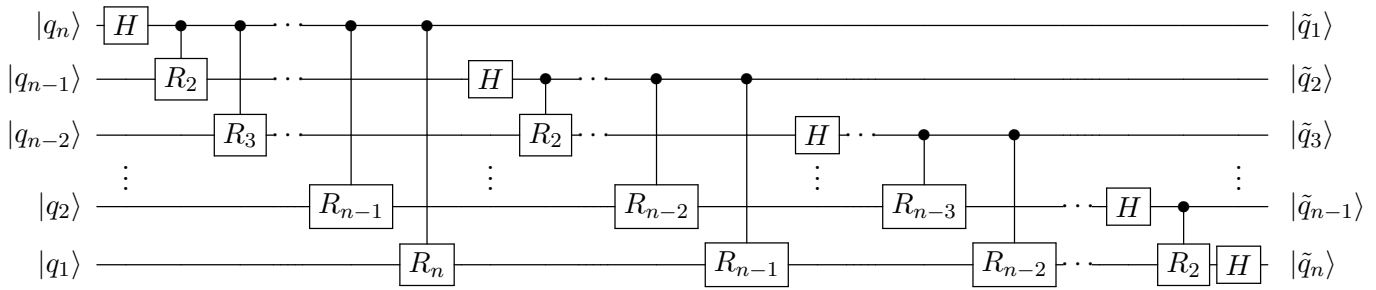

\noindent
It can be shown that the QFT is an unitary operator \cite{nielsen_chuang}.

The advantage of the QFT is that it can be performed with just $\mathcal{O}(n^2)$ operations (quantum gates). This is in sharp contrast to the classical fast Fourier transform (FFT) with the scaling $\mathcal{O}(N\mathrm{log}N=n2^n)$. Without the derivation, which can be found e.g. in \cite{nielsen_chuang}, the quantum circuit for the QFT is shown in Figure \ref{qft}. $H$ (Hadamard) and $R_j$ gates are represented by the following matrices

\begin{equation}
  H = \frac{1}{\sqrt{2}}\left(\begin{array}{rr} 1 & 1 \\ 1 & -1 \end{array}\right), \qquad R_j = \left(\begin{array}{rr} 1 & 0 \\ 0 & e^{2\pi i/2^j} \end{array}\right).
\end{equation}

It should be noted that even though the QFT can be done \textit{exponentially} faster than the FFT, it cannot be used as an \textit{efficient} straightforward replacement of the Fourier transform itself. It would indeed require to prepare an arbitrary state of $n$ qubits and also measure all of the complex amplitudes at the end, which cannot be done \textit{efficiently}. Nevertheless the QFT is a key part of the phase estimation algorithm \cite{nielsen_chuang} (contained also in the Shor's factoring algorithm \cite{shor_1994, shor_1997}) as will be shown in Section \ref{pea_section}.

\subsection{Semiclassical approach to quantum Fourier transform}
\label{mqft}
The QFT circuit from Figure \ref{qft} can in fact be greatly simplified using the semiclassical (measurement based) approach \cite{griffiths_1996}. Notice that gates acting on each qubit [except the first (top most) and the last ones where the corresponding parts are missing] obey the general structure: first, $R_j$ gates controlled by previous qubits are applied, then the Hadamard gate is applied, and finally they serve as control qubits for the subsequent ones. Because the state of each qubit does not change after the application of the Hadamard gate, we can in fact do the measurement just after this gate. Instead of employing controlled $R_j$ gates, we then apply only the corresponding one qubit gates depending on the results of individual measurements. Moreover, all $R_j$ gates acting on a $k$th qubit can be merged into a single rotation gate

\begin{equation}
  R_z(\omega_k) = \left(\begin{array}{rr} 1 & 0 \\ 0 & e^{2\pi i \omega_k} \end{array}\right),  
\end{equation}

\noindent
whose angle $\omega_k$ depends on the results of previously measured qubits ($q_i$) according to the formula

\begin{equation}
  \omega_k = \sum_{i=2}^{n-k+1} \frac{q_{k+i-1}}{2^i}, \qquad k:~n \longrightarrow 1.
  \label{omega}
\end{equation}

\begin{figure}[!ht]
  \mbox{
    \Qcircuit @C=0.8em @R=0.8em {
      \lstick{\ket{q_k}} & \gate{R_{z}(\omega_{k})} & \gate{H} & \meter & \cw & q_k
    }
  }
  \caption{Simplified, measurement based circuit for the $k$th qubit of the QFT.}
  \label{iqft}
\end{figure}

Figure \ref{iqft} shows the semiclassical QFT circuit pattern which is the same for all qubits. The big advantage of the aforementioned approach is that we have actually replaced two qubit gates by one qubit ones (controlled by a classical signal). This technique is especially useful in connection with the phase estimation algorithm where it leads to the formulation of its iterative version (IPEA, Section \ref{ipea}).

\subsection{Phase estimation algorithm}
\label{pea_section}
The phase estimation algorithm (PEA) \cite{nielsen_chuang} is a quantum algorithm for obtaining the eigenvalue of an unitary operator $\hat{U}$, based on a given initial guess of the corresponding eigenvector. Since an unitary $\hat{U}$ can be written as $\hat{U} = e^{i\hat{H}}$, with $\hat{H}$ Hermitian, the PEA can be viewed as a quantum substitute of the classical diagonalization.

Suppose that $\ket{u}$ is an eigenvector of $\hat{U}$ and that it holds

\begin{equation}
  \hat{U} | u \rangle = e^{2\pi i \phi} | u \rangle, \qquad \phi \in \langle 0, 1),
\end{equation}

\noindent
where $\phi$ is the phase which is estimated by the algorithm. Quantum register is divided into two parts. The first one, called the read-out part, is composed of $m$ qubits on which the binary representation of $\phi$ is measured at the end and which is initialized to the state $| 0 \rangle^{\otimes m}$. The second part contains the corresponding eigenvector $| u \rangle$. 

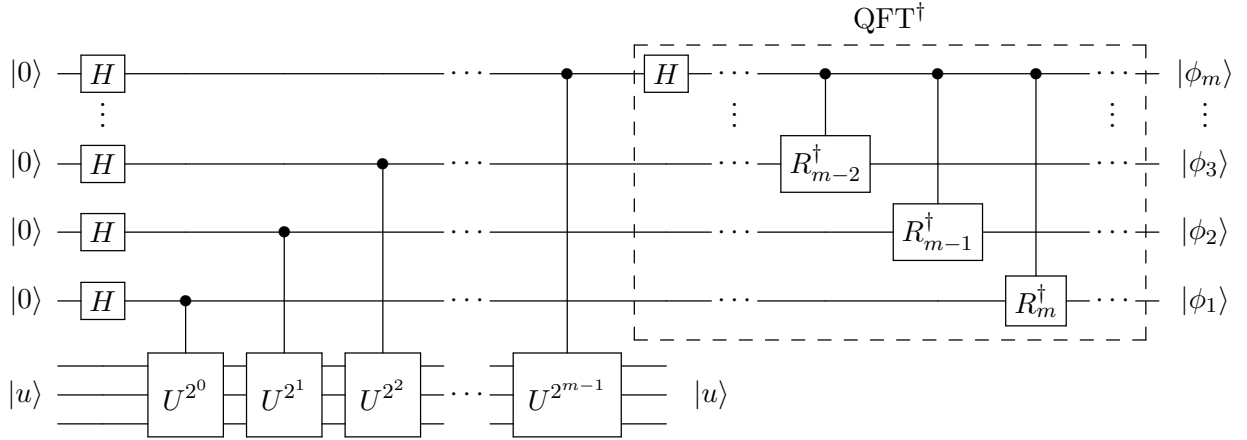
\begin{figure}[!ht]
\begin{center}
  \hskip 7cm
  QFT$^{\dagger}$ \\[0.1cm]
  \mbox{ 
    \Qcircuit @C=0.8em @R=0.1em {
      \lstick{\ket{0}} & \gate{H} & \qw & \qw & \qw & \qw & \cdots & & \ctrl{31} & \gate{H} & \qw & \cdots & & \ctrl{15} & \ctrl{19} & \ctrl{23} & \qw & \cdots & & \qw & & \ket{\phi_m} \\ \\ \\ \\ \\ 
       & \vdots & & & & & & & & & & \vdots & & & & & & \vdots & & & & \vdots \\ \\ \\ \\ \\ \\ \\ \\ \\ \\
      \lstick{\ket{0}} & \gate{H} & \qw & \qw & \ctrl{16} & \qw & \cdots & & \qw & \qw & \qw & \cdots & & \gate{R^{\dagger}_{m-2}} & \qw & \qw & \qw & \cdots & & \qw  & & \ket{\phi_3} \\ \\ \\ \\
      \lstick{\ket{0}} & \gate{H} & \qw & \ctrl{12} & \qw & \qw & \cdots & & \qw & \qw & \qw & \cdots & & \qw & \gate{R^{\dagger}_{m-1}} & \qw & \qw & \cdots & & \qw  & & \ket{\phi_2} \\ \\ \\ \\
      \lstick{\ket{0}} & \gate{H} & \ctrl{8} & \qw & \qw & \qw & \cdots & & \qw & \qw & \qw & \cdots & & \qw & \qw & \gate{R^{\dagger}_{m}} & \qw & \cdots & \push{\rule{0em}{2em}} & \qw  & & \ket{\phi_1} \\ \\ \\ \\ \\ \\ \\ \\
       & \qw & \multigate{2}{\displaystyle{U^{2^0}}} & \multigate{2}{\displaystyle{U^{2^1}}} & \multigate{2}{\displaystyle{U^{2^2}}} & \qw & & & \multigate{2}{\displaystyle{U^{2^{m-1}}}} & \qw \\
       \lstick{\ket{u}} & \qw & \ghost{\displaystyle{U^{2^0}}} & \ghost{\displaystyle{U^{2^1}}} & \ghost{\displaystyle{U^{2^2}}} & \qw & \cdots & & \ghost{\displaystyle{U^{2^{m-1}}}} & \qw & \ket{u} \\
       & \qw & \ghost{\displaystyle{U^{2^0}}} & \ghost{\displaystyle{U^{2^1}}} & \ghost{\displaystyle{U^{2^2}}} & \qw & & & \ghost{\displaystyle{U^{2^{m-1}}}} & \qw \gategroup{1}{10}{24}{19}{.7em}{--}
    }
  }
  \caption{The PEA circuit with the highlighted part corresponding to the inverse QFT.}
  \label{pea}
\end{center}
\end{figure}

The PEA quantum circuit is shown in Figure \ref{pea}. The application of Hadamard gates on all qubits in the first part of the register gives

\begin{equation}
 | \mathrm{reg} \rangle = \frac{1}{\sqrt{2^{m}}} \sum_{j=0}^{2^{m}-1} | j \rangle | u \rangle .
 \label{after_h}
\end{equation}

\noindent
Next, after the application of a sequence of controlled powers of $\hat{U}$, the register is transformed into

\begin{equation}
| \mathrm{reg} \rangle = \frac{1}{\sqrt{2^{m}}} \sum_{j=0}^{2^{m}-1} \hat{U}^{j} | j \rangle | u \rangle = \frac{1}{\sqrt{2^{m}}} \sum_{j=0}^{2^{m}-1} e^{2\pi ij \phi} | j \rangle | u \rangle. 
 \label{after_cu} 
\end{equation}

\noindent
The heart of the PEA is the inverse quantum Fourier transform (QFT$^{\dagger}$, highlighted in Figure \ref{pea}) performed on the read-out part of the register which is transformed to $| 2^{m}\phi \rangle | u \rangle$. 

If the phase can be expressed exactly in $m$ bits

\begin{equation}
 \phi = 0.\phi_{1}\phi_{2} \ldots \phi_{m} 
  = \frac{\phi_{1}}{2} + \frac{\phi_{2}}{2^{2}} + \ldots + \frac{\phi_{m}}{2^{m}}, \qquad \phi_{i} \in \{0,1\}, 
\end{equation}

\noindent
it (and consequently the eigenvalue) is recovered with unity probability by a measurement on the first part of the quantum register.

\begin{figure}[!t]
  \begin{center}
    \includegraphics[width=10cm]{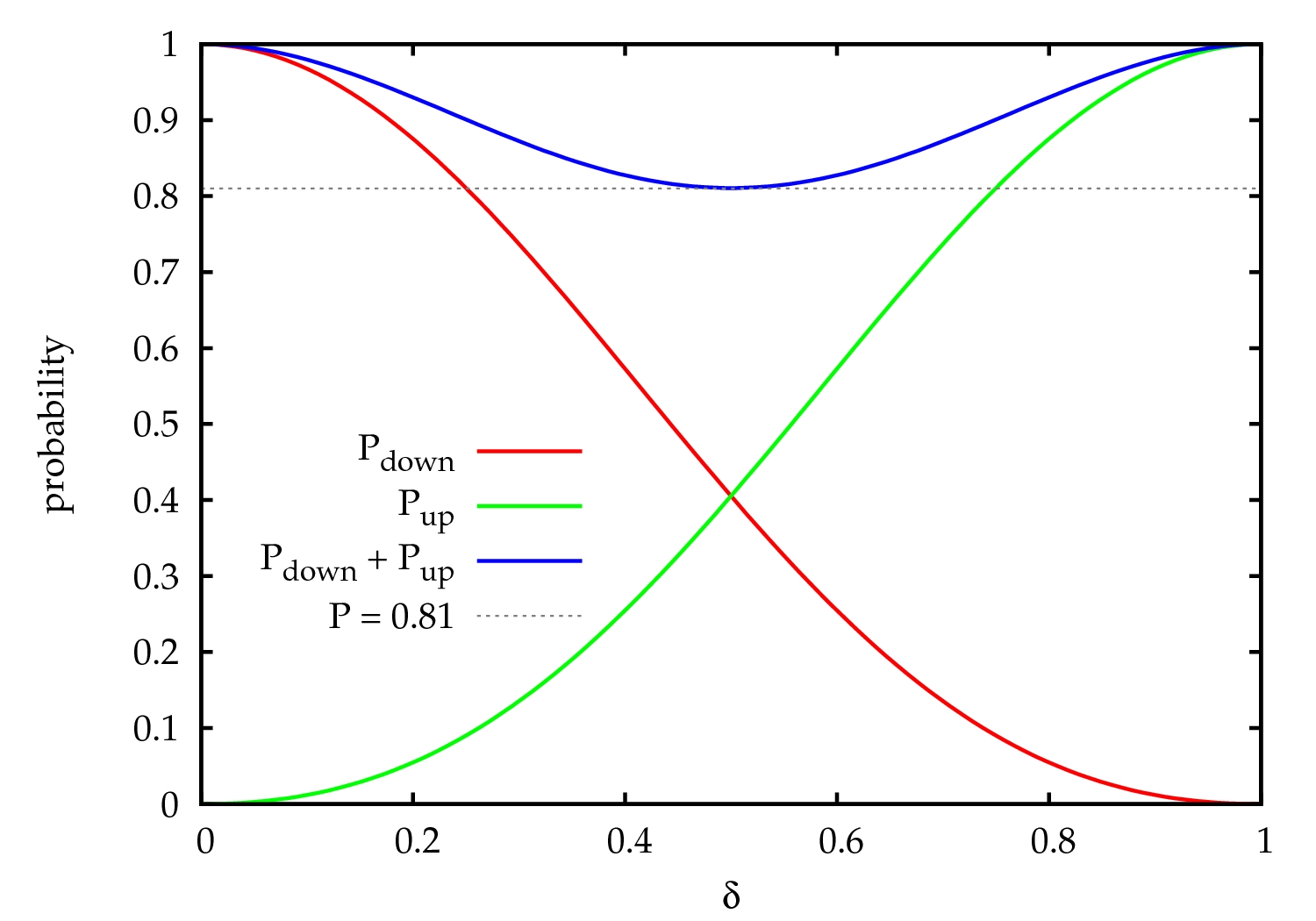}
    \caption{The dependence of success probabilities of the PEA on $\delta$ for $m=20$. $P_{\rm{down}}$ and $P_{\rm{up}}$ denote the success probabilities corresponding to rounding the exact phase up/down to $m$ binary digits.}
    \label{pea_prob}
  \end{center}
\end{figure}

The situation is more complicated when $\phi$ cannot be expressed exactly in $m$ bits. Then we can write

\begin{equation}
 \phi = \tilde{\phi} + \delta 2^{-m},  
 \label{reminder}
\end{equation}

\noindent
where $\tilde{\phi} = \phi_{1}\phi_{2}\ldots\phi_{m}$ denotes the first $m$ bits of the binary expansion and $\delta:~0 \le \delta < 1$ is a remainder. The closest $m$-bit estimators of $\phi$ correspond to either $\tilde{\phi}$ (rounding down) or $\tilde{\phi} + 2^{-m}$ (rounding up). When we label the probabilities of measuring these two estimators by $P_{\rm{down}}$ and $P_{\rm{up}}$, it can be shown (e.g. \cite{dobsicek_phd}) that the sum $P_{\rm{down}} + P_{\rm{up}}$ decreases monotonically with increasing $m$. The dependence of $P_{\rm{down}}$ and $P_{\rm{up}}$ on $\delta$ for $m = 20$ is presented in Figure \ref{pea_prob}. In the limit $m \rightarrow \infty$, the lower bound reads \cite{dobsicek_phd}

\begin{equation}
 P_{\rm{down}}(\delta = 1/2) + P_{\rm{up}}(\delta = 1/2) = \frac{4}{\pi^{2}} + \frac{4}{\pi^{2}} > 0.81.
 \label{sc_factor}
\end{equation} 

If the desired eigenvector is not known explicitly (as is typically the case in quantum chemistry), we can start the algorithm with an arbitrary initial guess vector $| \psi \rangle$, which can be expanded in terms of eigenvectors of $\hat{U}$

\begin{equation}
 | \psi \rangle = \sum_{i} c_{i} | u_{i} \rangle. 
\end{equation}

\noindent
The probability of obtaining the \textit{exact} $m$-bit $\phi_{i}$ is due to linearity of the algorithm $|c_{i}|^{2}$. It is important to note that the initial guess does not influence the accuracy of the phase, only the probability with which the phase of a particular eigenstate is measured. When $\phi_i$ cannot be expressed in $m$ bits, the lower bound for $P_{\rm{down}} + P_{\rm{up}}$ corresponding to $\phi_i$ is equal to $0.81\cdot|c_{i}|^{2}$.

\subsection{Iterative phase estimation algorithm}
\label{ipea}
\begin{figure}[!h]
 \begin{center}  
 \mbox{
   \Qcircuit @C=1em @R=1em {
     \lstick{\ket{0}} & \gate{H} & \ctrl{1} & \gate{R_{z}(\omega_{k})} & \gate{H} & \meter & \cw & \phi_{k} \\
     \lstick{\ket{u}} & {/} \qw & \gate{U^{2^{k-1}}} & {/} \qw & \qw
   }
 }
 \end{center}
 \caption{The $k$-th iteration of the IPEA. The feedback angle
depends on the previously measured bits, $k$ is iterated backwards from $m$ to 1.}
 \label{ipea_iteration} 
\end{figure}
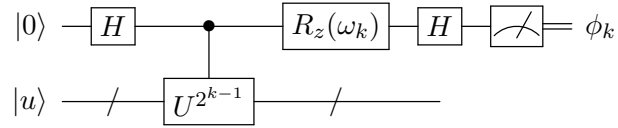

Using the semiclassical QFT \cite{griffiths_1996} (Section \ref{mqft}), the PEA circuit can be simplified, having only one ancillary qubit in the read-out part of the quantum register. The algorithm then proceeds in an iterative manner [iterative phase estimation algorithm (IPEA)]. The $k$-th iteration of this scheme is presented in Figure \ref{ipea_iteration}. Note that as the PEA uses the inverse QFT, the angle $\omega_k$ (\ref{omega}) must be negative now. The algorithm is iterated backward from the least significant bits of $\phi$ to the most significant ones, for $k$ going from $m$ to 1. For our purposes, the presented IPEA, which is the \textit{unitary} matrix eigenvalue algorithm, is completely adequate, but we would like to note that Wang et al. recently presented a modified version of the IPEA capable of finding eigenvalues of \textit{non-unitary} matrices \cite{wang_2010}.

The IPEA is in fact completely equivalent to the original (multiqubit) PEA \cite{dobsicek_phd}. It exhibits the same behaviour - decreasing of the success probability when the phase cannot be expressed exactly in a particular number of bits. One possibility of a success probability amplification is performing more iteration steps (more than is the desired accuracy of $\phi$): when extracting $m' = m + \log (2 + 1/2 \epsilon)$ bits, the phase is accurate to $m$ binary digits with probability at least $1 - \epsilon$ \cite{nielsen_chuang}. This method is however not very useful, since implementing $\hat{U}^{2^{k-1}}$ for large $k$ is the algorithm's bottleneck in a realistic noisy environment \cite{dobsicek_2007}.  

Another alternative \cite{dobsicek_2007} is to repeat the measurement for the least important bits of the phase binary expansion. Using the majority voting (for bit value 0 or 1), the effective error probability decreases exponentially with the number of repetitions according to the binomial distribution. This measurement repetition only for the few least important bits of $\phi$ is unfortunately possible only if the exact eigenstates of $\hat{U}$ are available.

\begin{figure}[!h]
  \begin{minipage}{\linewidth}
  \centering
  \mbox{
    \Qcircuit @C=0.7em @R=0.7em {
      & \lstick{\ket{0}} & \gate{H} & \ctrl{1} & \gate{R_{z}(\omega_{m})} & \gate{H} & \meter & \cw & ~x_{m} & & & & &  \lstick{\ket{0}} & \gate{H} & \ctrl{1} & \gate{R_{z}(\omega_{m-1})} & \gate{H} & \meter & \cw & & x_{m-1} \\
      & \gate{\rm{ISP}} & {/} \qw & \gate{U^{2^{m-1}}} & \qw & \qw & \qw & {/} \qw & \qw & \qw & \qw & \qw & \qw & \qw & \qw & \gate{U^{2^{m-2}}} & {/} \qw & \qw
    }
  }
  \vskip 0.3cm  
  a) Maintaining the second part of the quantum register during all iterations (version \textbf{A}).
  \end{minipage}

  \vskip 0.5cm
  
  \begin{minipage}{\linewidth}
  \centering
  \mbox{
    \Qcircuit @C=0.7em @R=1em {
      & \lstick{\ket{0}} & \gate{H} & \ctrl{1} & \gate{R_{z}(\omega_{m})} & \gate{H} & \meter & \cw & ~x_{m} & & & &  \lstick{\ket{0}} & \gate{H} & \ctrl{1} & \gate{R_{z}(\omega_{m-1})} & \gate{H} & \meter & \cw & & x_{m-1} \\
      & \gate{\rm{ISP}} & {/} \qw & \gate{U^{2^{m-1}}} & {/} \qw & \qw & & & & & & & \gate{\rm{ISP}} & {/} \qw & \gate{U^{2^{m-2}}} & {/} \qw & \qw
    }  
  }  
  \vskip 0.3cm    
  b) Repeated initial state preparation in each iteration (version \textbf{B}).
  \end{minipage}  
  \caption{Comparison of the two versions of IPEA, \fbox{ISP} denotes the initial state preparation.}
  \label{ipea_comparison}
\end{figure}
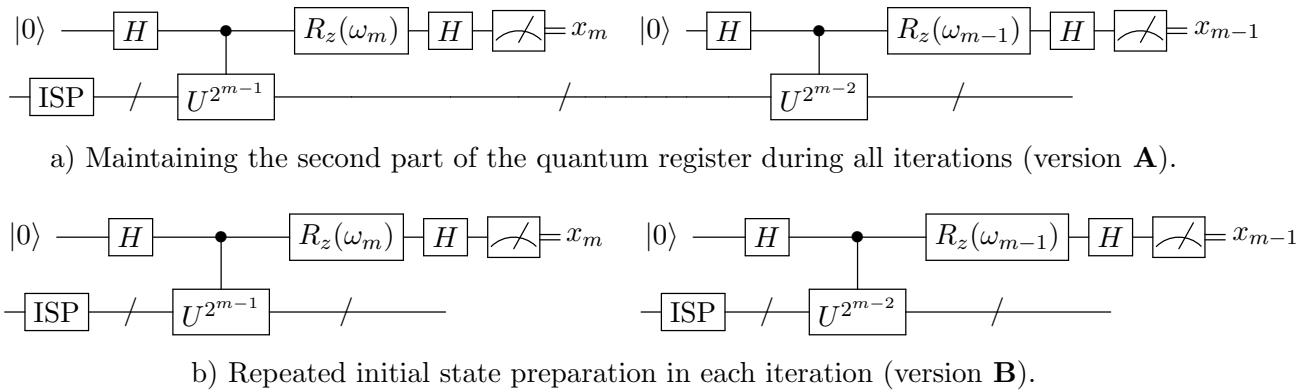

\begin{figure}[!ht]
  \begin{center}
    \includegraphics[width=10cm]{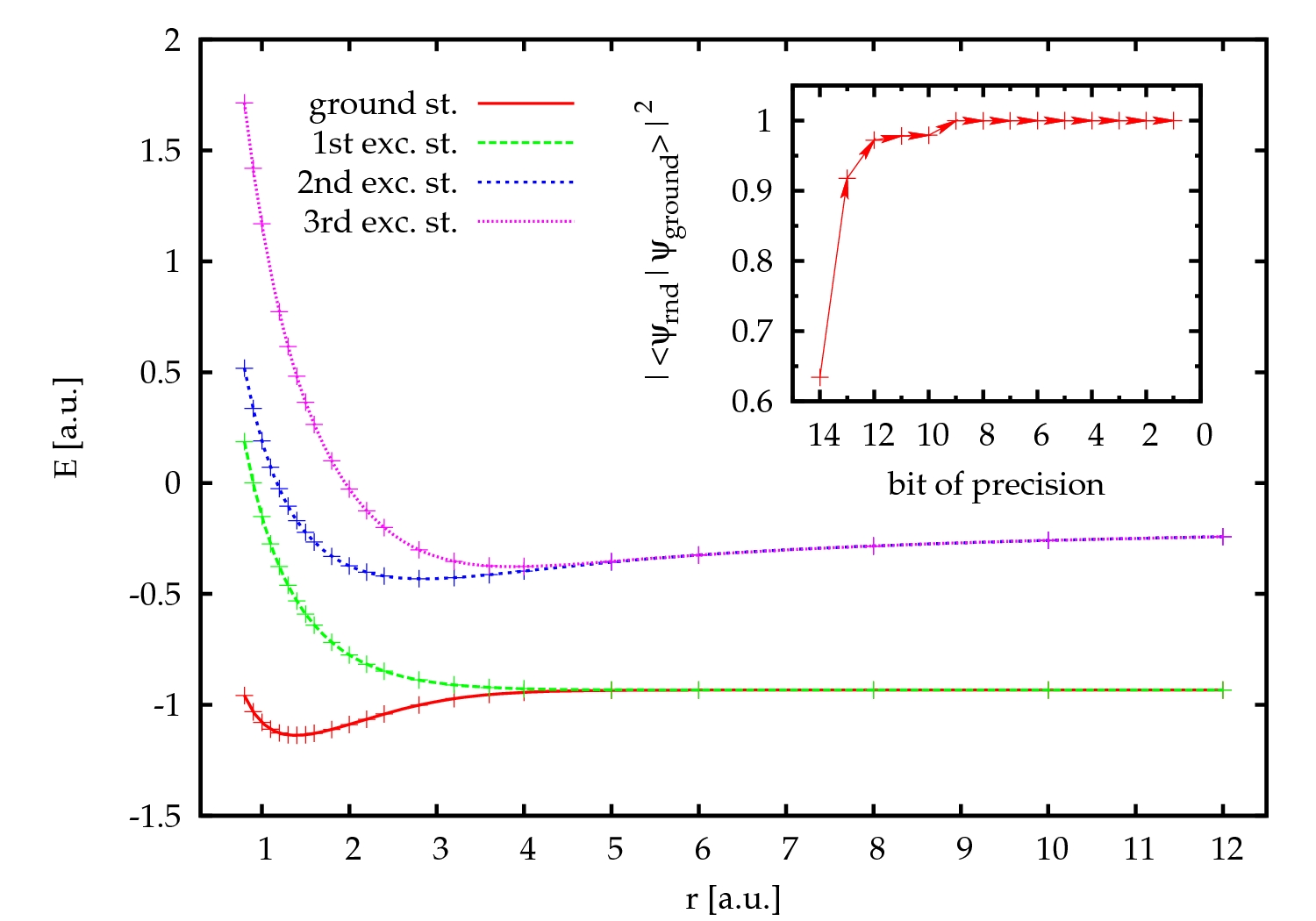}
    \caption{Energies of the four electronic states of H$_{2}$ in STO-3G basis which were obtained by the qFCI method (IPEA version \textbf{A}) with randomly generated initial guess states. Small figure inside presents the increasing overlap between the actual state of the second part of the quantum register and the exact wave function for one of random runs of the algorithm leading to the ground state. Figure reprinted from \cite{veis_2010}.}
    \label{h2_rnd}
  \end{center} 
\end{figure}

When working with general initial states (as in a practical application to quantum chemistry), two scenarios  are possible \cite{veis_2010}, as shown in Figure \ref{ipea_comparison}. Maintaining the second part of the quantum register during all iterations and amplification of the success probabilities by repeating the whole process number of times is the first possibility. This version was denoted as \textbf{A} version of IPEA. The biggest advantage of this approach is that one always ends up with one of the eigenstates of $\hat{U}$ in the second part of the quantum register as was also the case of the PEA. It happens through successive collapses of the system state into the corresponding eigensubspace and is demonstrated on the hydrogen molecule with random initial states in Figure \ref{h2_rnd}. The biggest disadvantage that complicates the potential physical realization of this scheme is the requirement for a long coherence time of the quantum register. We would like to note here that when amplifying the success probability by repeating the whole process, it must be higher than 0.5 to be sure that we get the energy of the right state. This, however, is not necessary for the ground state energy which can always be identified by the lowest eigenvalue \cite{veis_2010,li_2011}.

Another possibility is to initialize the second part of the quantum register at every iteration step (\textbf{B} version of IPEA). Every iteration step (not only the least important bits of $\phi$ as in \cite{dobsicek_2007}) must be repeated and measurement statistics performed. One could otherwise possibly mix bits belonging to different eigenvalues in different iterations and obtain an unphysical result. The biggest advantage of this approach is avoidance of the long coherence times and therefore potentially easier physical implementation. On the other hand, the biggest disadvantage is that no improving of the overlap between the actual state of the quantum register and the exact wave function occurs during the iterations and one must ``fight" the overlap problem at every iteration step. But as will be shown in Section \ref{NR} on the example of the methylene molecule \cite{veis_2010}, small number of repetitions of each iteration is sufficient for amplification of the success probability to unity, when a suitable initial state of the quantum register is used.

At the end of this section, we would like to mention a different way of reducing the number of read-out qubits required by the PEA, which was suggested in the seminal work by Aspuru-Guzik et al. \cite{aspuru-guzik_2005}. Their recursive variant of the PEA uses four qubits in the read-out part of the quantum register on which the phase (and therefore also the energy) is successively improved. It starts with measuring the first four bits of the phase $\phi$. The Hamiltonian is then shifted by this reference value and a four-bit estimate of the deviation of the phase from the reference one on the half of the interval computed. The procedure is iteratively repeated and the overall effect is a gain of one additional bit of $\phi$ at each iteration step (thus the same as in the IPEA). In spite of the fact that this method uses four read-out qubits instead of one which is used by the IPEA, it is worth mentioning. First of all, it was the first iterative version of the PEA applied to the Hamiltonian eigenvalue problem \cite{aspuru-guzik_2005}. Secondly, it recovers the energy starting from the most important bits towards the least important ones (in contrast to the IPEA), which can be advantageous in certain situations.

\section{Quantum full configuration interaction method}
\label{Quantum full configuration interaction method}
The PEA/IPEA can in fact be used for efficient computations of eigenvalues of local Hamiltonians \cite{abrams_1999}. If we take $\hat{U}$ in the form

\begin{equation}
  \hat{U} = e^{i\tau\hat{H}},  
\end{equation}

\noindent
where $\hat{H}$ is a local Hamiltonian and $\tau$ a suitable parameter assuring $\phi$ being in the interval $\langle 0,1)$, then the algorithm reveals the energy spectrum of $\hat{H}$. The whole procedure can be simply viewed as a time propagation of a trial wave function followed by the QFT switching from the time to energy domain and a measurement projecting out a certain eigenstate.

In this section, we discuss the application to molecular Born-Oppenheimer electronic Hamiltonians. We will start with a mapping of quantum chemical wave functions onto a quantum register (Section \ref{mapping}). Section \ref{init_states} briefly mentions the initial state preparation and Section \ref{c_U} deals with the crucial part of the algorithm, namely the implementation of controlled powers of the exponential of molecular Hamiltonians (controlled ``time propagation").

\subsection{Mapping of quantum chemical wave functions onto quantum register}
\label{mapping}
Several possible mappings of a quantum chemical wave function onto a register of qubits have been proposed. The simplest, but the least economical one in terms of the number of employed qubits, is so called direct mapping \cite{aspuru-guzik_2005}. In this approach, individual spin orbitals are directly assigned to qubits, since each spin orbital can be either occupied or unoccupied, corresponding to $|1\rangle$ or $|0\rangle$ states. The inefficiency lies in the fact that it actually maps the whole Fock space of the system (states with different number of electrons) on the Hilbert space of the quantum register. Relativistic generalization of this approach \cite{veis_2011} assigns one qubit to one Kramers pair bispinor ($A$ or $B$, analogy to $\alpha$ and $\beta$ spin in non-relativistic treatment). The advantage of the direct mapping stems from the fact, that a general factorization scheme, i.e. an algorithm to systematically generate a quantum circuit implementing the exponential of a Hamiltonian has been discovered (see Section \ref{fact}).

Compact mappings from a subspace of fixed-electron-number wave functions, spin-adapted \cite{aspuru-guzik_2005}, or even symmetry-adapted wave functions employing the point group \cite{wang_2008} or double group symmetry \cite{veis_2011} to the register of qubits have also been proposed. However, general factorization schemes are unfortunately not known for these mappings. The factorization to elementary quantum gates can be for small circuits performed either with numerical optimization  techniques (e.g. with genetic algorithms \cite{daskin_2011}) or analytically \cite{shende_2006}, but neither \textit{efficiently}. Its use is motivated by the need to employ as few qubits as possible in today's experimental realizations \cite{lanyon_2010,du_2010}.

\subsection{Initial states for the algorithm}
\label{init_states}
The quantum full configuration interaction (qFCI) algorithm must be started with some initial guess state. Generally, it holds that the closer is the initial guess to the exact wave function corresponding to the calculated energy, the higher is the success probability of measuring the energy. As was shown in \cite{wang_2009, veis_2010}, the simplest one-determinantal Hartree-Fock guess may not be successful in situations, where correlation (particularly the static one) plays an important role. In these situations, initial guess states from more sophisticated \textit{polynomially} scaling methods can be used [e.g. complete active space self consistent field (CASSCF) method in a limited orbital CAS]. 

Preparing a general initial state (an arbitrarily chosen vector from the Hilbert space of $n$ qubits) is a hard task as this vector can contain up to $2^{n}$ non-zero components in general and it cannot be performed \textit{efficiently}. Fortunately, initial guesses including only few determinants in a superposition are sufficient for most purposes of quantum chemistry \cite{veis_2010}. These states can be prepared e.g. with the procedure described by Ortiz et al. \cite{ortiz_2001} which scales as $\mathcal{O}(N^{2})$ in the number of determinants $N$. Preparation of general molecular-like states from the combinatorial space of dimension $\big(\begin{array}{c} n \\ m \end{array}\big)$ corresponding to distributing $m$ electrons among $n$ spin orbitals was presented in \cite{wang_2009}. Preparation of many-particle states in a superposition on a lattice which can be then propagated by quantum chemical dynamics algorithm \cite{kassal_2008} was studied in \cite{ward_2009}.

\subsubsection{Adiabatic state preparation}
A different approach of the initial state preparation is the adiabatic state preparation (ASP) method of Aspuru-Guzik et al. \cite{aspuru-guzik_2005}. In the ASP method, one slowly varies the Hamiltonian of the quantum register, starting with a trivial one and the register in its (exactly known) ground state and ending with the final exact one in the following simple way

\begin{equation}
  \hat{H} = (1-s)\hat{H}_{\rm{init}} + s\hat{H}_{\rm{exact}} \qquad s: 0 \longrightarrow 1.
\end{equation} 

\noindent
If the change is slow enough (depending on the gap between the ground and the first excited state), the register remains in its ground state according to the adiabatic theorem \cite{fahri_science_2001}. \mbox{In the compact} mapping, $\hat{H}_{\rm{init}}$ can be defined to have all matrix elements equal to zero, except $H_{11}$, which is equal to the (Dirac-)Hartree-Fock energy \cite{aspuru-guzik_2005,veis_2011}. 

\begin{figure}[!ht]
  \begin{center}
    \subfloat[][]{  
      \includegraphics[width=7.8cm]{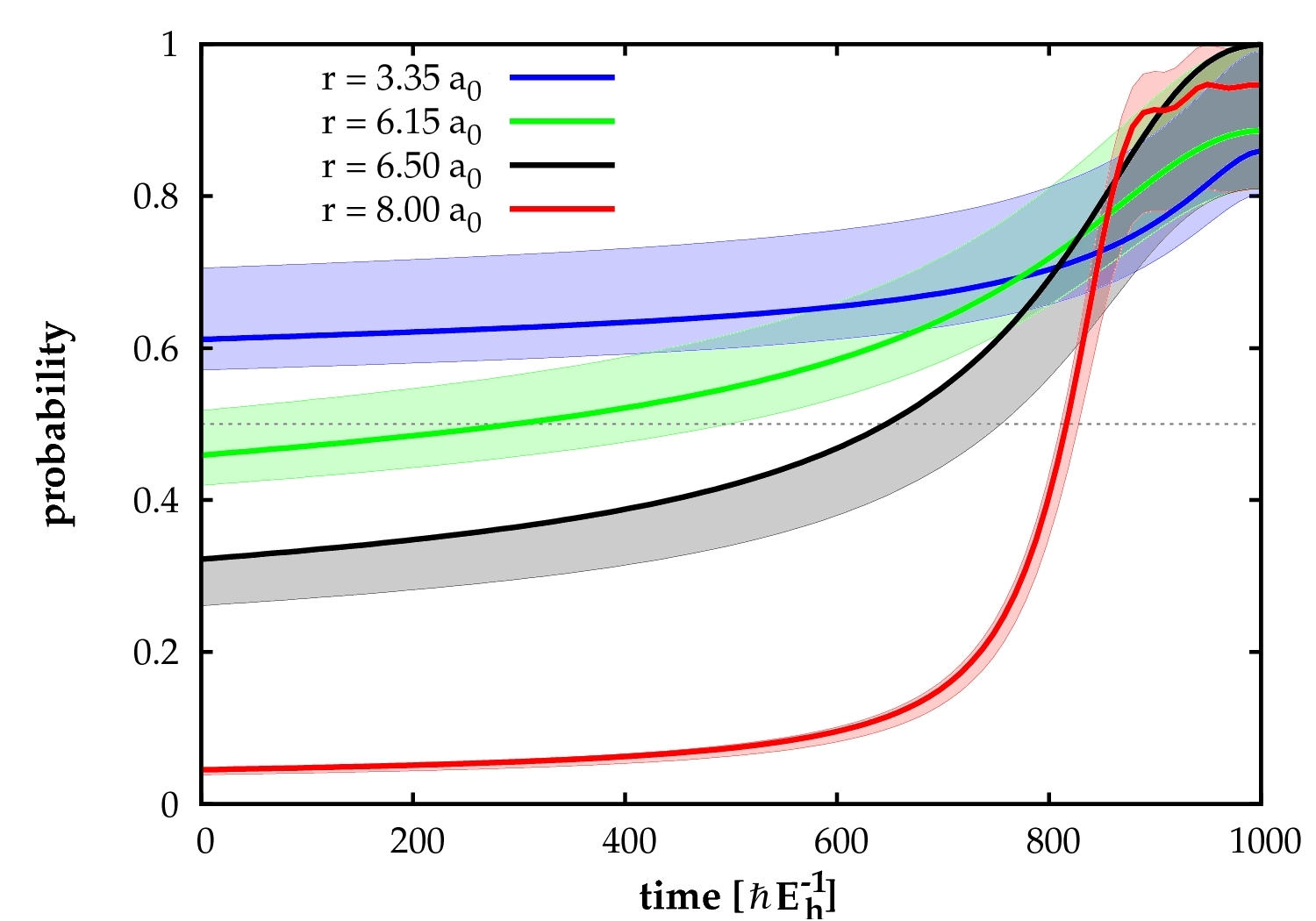}
      \label{aspA}
    } 
    \subfloat[][]{
      \includegraphics[width=7.8cm]{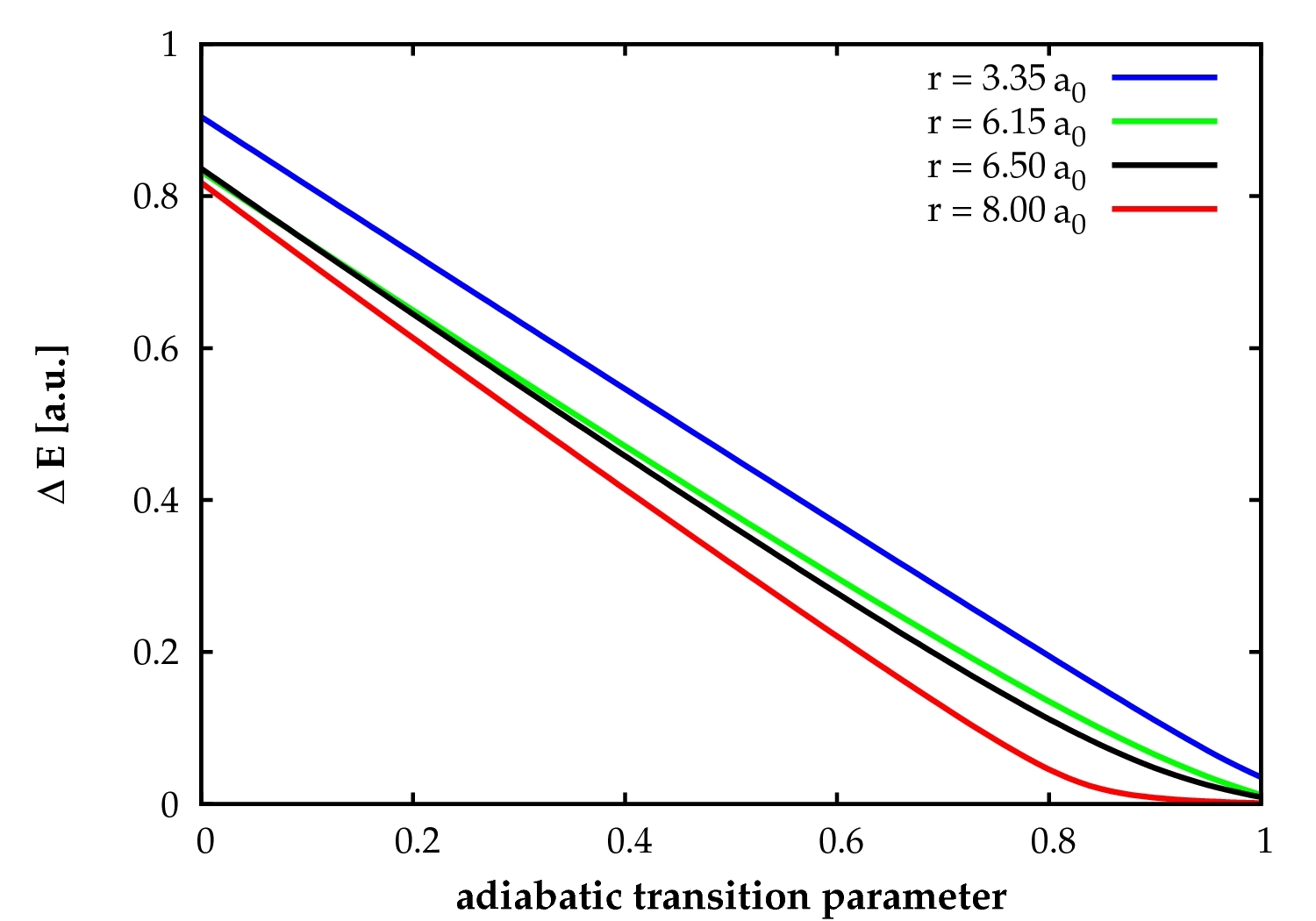}
      \label{aspB}
    }   
    \caption{Adiabatic state preparation (ASP) of the SbH ground state for different internuclear distances. (a) Dependence of the IPEA success probability on time during the ASP (1000 $\hbar E_{h}^{-1}$ $\approx 10^{-14}$ s). Solid lines correspond to the success probabilities, $|\langle \psi_{\rm{ASP}} | \psi_{\rm{exact}} \rangle |^{2} \cdot (0.81,1\rangle$ interval is coloured. Reprinted from \cite{veis_2011}. (b) Dependence of the energy gap between the ground and the first excited state on the adiabatic transition parameter $s$.}
  \label{asp}
  \end{center}
\end{figure}

Figure \ref{aspA} demonstrates on the example of the SbH molecule the improvement of the IPEA (version \textbf{A}) ground state success probability during the ASP procedure. The dependence of the energy gap $(\Delta E)$ between the ground and the first excited state on the adiabatic transition parameter $s$ is shown in Figure \ref{aspB}. Although $\Delta E$ is getting close to 0 for \mbox{$r = 8.0$ $a_0$} and $s$ going to 1, the ground state is becoming degenerate at this internuclear distance and this fact thus does not influence the IPEA success probability. We will discuss the relativistic qFCI method and its application to the SbH spin-orbit splitting in more detail in Section \ref{R}.

Recently, the ASP of the hydrogen molecule ground state has been realized experimentally on a NMR quantum simulator \cite{du_2010}.

\subsection{Controlled ``time propagation"}
\label{c_U}
To study the overall scaling of the qFCI algorithm, one must decompose the only multi qubit gate from Figure \ref{ipea_iteration}, i.e. controlled powers of $\hat{U} = e^{i\tau \hat{H}}$ to elementary one and two qubit gates.

For this purpose, it is convenient to express the electronic Hamiltonian in the second quantized form \cite{szabo_ostlund}

\begin{equation}
 \hat{H} = \sum_{pq} h_{pq} \hat{a}_{p}^{\dagger} \hat{a}_{q} + \frac{1}{2} \sum_{pqrs} g_{pqrs} \hat{a}_{p}^{\dagger} \hat{a}_{q}^{\dagger} \hat{a}_{s} \hat{a}_{r} = \sum_{X=1}^{L} \hat{h}_{X},
  \label{ham_sec_quant}
\end{equation}

\noindent
where $h_{pq}$ and $g_{pqrs}$ are one- and two-electron integrals and $\hat{a}_{p}^{\dagger}$ and $\hat{a}_{p}$ are fermionic creation and annihilation operators. The whole summation is formally expressed as a sum of individual terms $\hat{h}_{X}$. The molecular integrals $h_{pq}$ and $g_{pqrs}$ can be \textit{efficiently} precalculated on a conventional computer \cite{helgaker} and represent a classical input to the quantum algorithm. In non-relativistic case, they are real-valued, while in relativistic case, they are in general complex.

Since the creation and annihilation operators generally do not commute, the exponential of a Hamiltonian cannot be written as a product of the exponentials of individual $\hat{h}_{X}$, but a numerical approximation must be used \cite{lloyd_1996}. The first-order Trotter approximation \cite{trotter} can be expressed as

\begin{equation}
 e^{i\tau\hat{H}} = e^{i\tau\sum_{X=1}^{L}\hat{h}_{X}} = \Big(\prod_{X=1}^{L} e^{i\hat{h}_{X}\tau/N}\Big)^{N} + \mathcal{O}(\tau^{2}/N).
 \label{trotter}
\end{equation}

\noindent
By choosing $N \geq (\tau^{2}/\epsilon)$, we can implement $\hat{U}$ within an error tolerance of $\mathcal{O}(\epsilon)$ using $\mathcal{O}(L(\tau^{2}/\epsilon))$ particular terms $e^{i\hat{h}_{X}\tau/N}$. 

Before discussing the factorization of these terms to elementary quantum gates in Section \ref{fact}, we would like to mention the implementation modification we use \cite{veis_2010}. Two more external inputs are necessary in our case. These are maximum ($E_{\mathrm{max}}$) and minimum ($E_{\mathrm{min}}$) energies expected in the studied system and we use $\hat{U}$ in the form

\begin{equation}
 \hat{U} = e^{i\tau(E_{\mathrm{max}} - \hat{H})}.
 \label{our_u}
\end{equation}

\noindent
For $\tau$, it holds

\begin{equation}
 \tau = \frac{2\pi}{E_{\mathrm{max}} - E_{\mathrm{min}}}
\end{equation}

\noindent
and the final energy is obtained according to the formula

\begin{equation}
 E = E_{\mathrm{max}}-\frac{2\pi\phi}{\tau}.
\end{equation}

\noindent
The modification mentioned above assures $\phi$ to be in the interval $\langle 0,1)$. 

$E_{\mathrm{min}}$ and $E_{\mathrm{max}}$ can in fact be chosen arbitrarily, but one must be sure that the calculated energy is within this interval, otherwise one would end up with a nonphysical energy, due to the periodicity of $e^{2\pi i \phi}$. The maximum energy can be e.g. the upper bound provided by any classical variational (\textit{polynomially} scaling) method, techniques for calculation of lower bounds \cite{bazley_1960,lowdin_1962,lowdin_1965} can on the other hand give the minimum energy. The smaller the interval between them is, the less iterations of IPEA are necessary for desired precision of $E$. 

Taking $\hat{U}$ in the form (\ref{our_u}) does not pose any difficulties indeed and as the following circuit equality shows, just one more one-qubit rotation is needed.

\begin{center}
\mbox{
 \Qcircuit @C=1em @R=0.5em {
   & \ctrl{2} & \qw & & & \ctrl{2} & \gate{\left(\begin{array}{cc} 
                                         1 & 0 \\
                                         0 & e^{i\tau E_{\mathrm{max}}}
                                         \end{array}\right)} & \qw \\
   & & & = & & \\                                      
   & \gate{e^{i\tau (E_{\mathrm{max}} - \hat{H})}} & \qw & & & \gate{e^{-i\tau \hat{H}}} & \qw & \qw
 }
}
\end{center}

\subsubsection{Decomposition of unitary propagator to elementary quantum gates}
\label{fact}
The decomposition of the unitary propagator $e^{i\tau\hat{H}}$ to elementary quantum gates \cite{ortiz_2001,ovrum_2007,whitfield_2010} proceeds in the following manner. First, the Jordan-Wigner transform \cite{jordan_1928} is used to express the fermionic second quantized operators in terms of Pauli $\sigma$ matrices. The Jordan-Wigner transform has the form

\begin{equation}
 \hat{a}_{n}^{\dagger} = \Bigg( \bigotimes_{j=1}^{n-1} \sigma_{z}^{j} \Bigg) \otimes \sigma_{-}^{n}, \quad \hat{a}_{n} = \Bigg( \bigotimes_{j=1}^{n-1} \sigma_{z}^{j} \Bigg) \otimes \sigma_{+}^{n},
\end{equation}

\noindent
where $\sigma_{\pm} = 1/2(\sigma_{x} \pm i \sigma_{y})$ and the superscript denotes the qubit on which the matrix operates. The Hamiltonian (\ref{ham_sec_quant}) then can be rewritten using strings of $\sigma$ matrices. Finally, the exponentials of these strings are build up from single-qubit gates and controlled NOT operations (CNOTs) \cite{nielsen_chuang}. 

We will demonstrate this approach on the one-electron part of the Hamiltonian (with complex-valued molecular integrals)

\begin{equation}
  \hat{H}_1 = \sum_{pq} h_{pq} \hat{a}^{\dagger}_{p} \hat{a}_{q} = \sum_{pp} h_{pp}\hat{a}^{\dagger}_{p} \hat{a}_{p} + \sum_{p>q} \big( h_{pq} \hat{a}^{\dagger}_{p} \hat{a}_{q} + h_{qp} \hat{a}^{\dagger}_{q} \hat{a}_{p} \big).
  \label{h1}
\end{equation}

\noindent
Employing the Jordan-Wigner transform, the diagonal terms can be written as

\begin{equation}
  h_{pp} a^{\dagger}_{p}a_{p} = \frac{h_{pp}^{\rm{R}}}{2} (\mathbf{1} - \sigma_{z}^{p}),  
\end{equation}

\noindent
where $h_{pp}^{\rm{R}}$ is the real part of $h_{pp}$ [$h_{pp}^{\rm{I}}$ (the imaginary part) is equal to zero due to the Hermicity of $\hat{H}$]. For the exponentials holds

\begin{equation}
  e^{i\hat{h}_{X}\tau/N} = e^{i h_{pp} a^{\dagger}_{p}a_{p} \tau/N} = \left(\begin{array}{cc} 1 & 0 \\ 0 & e^{i h_{pp}\tau/N} \end{array}\right)^{(p)}.
\end{equation}

\noindent
The superscript $(p)$ at the matrix denotes the qubit on which the one qubit gate operates.

Similarly, the off-diagonal terms read
\clearpage

\begin{equation*}
  h_{pq} a^{\dagger}_{p} a_{q} + h_{qp} a^{\dagger}_{q} a_{p} =
\end{equation*}
\vskip -0.8cm
\begin{eqnarray}
  & = & \frac{h_{pq}^{\rm{R}}}{2} \Big[ \sigma_{x}^{p} \otimes \big( \sigma_{z}^{p \rightarrow q} \big) \otimes \sigma_{x}^{q} + \sigma_{y}^{p} \otimes \big( \sigma_{z}^{p \rightarrow q} \big) \otimes \sigma_{y}^{q}\Big] + \nonumber \\
  & + & \frac{h_{pq}^{\rm{I}}}{2} \Big[ \sigma_{y}^{p} \otimes \big( \sigma_{z}^{p \rightarrow q} \big) \otimes \sigma_{x}^{q} - \sigma_{x}^{p} \otimes \big( \sigma_{z}^{p \rightarrow q} \big) \otimes \sigma_{y}^{q}\Big],
\label{off_diag}
\end{eqnarray}

\noindent
where $\sigma_{z}^{p \rightarrow q}$ represents the direct product

\begin{equation}
  \sigma_{z}^{p \rightarrow q} \equiv \sigma_{z}^{p+1} \otimes \sigma_{z}^{p+2} \otimes \ldots \otimes \sigma_{z}^{q-2} \otimes \sigma_{z}^{q-1}.
\end{equation}

\noindent
Note that (\ref{off_diag}) contains the four aforementioned strings of $\sigma$ matrices. 

The exponential of the string of $\sigma_z$ matrices $\mathrm{exp}[i\tau(\sigma_z \otimes \ldots \otimes \sigma_z)]$ is in fact diagonal in the computational basis with the phase shift $e^{\pm i\tau}$ on the diagonal. The sign of this phase shift depends on the parity of the corresponding basis state (``+" if the number of ones in the binary representation is even, ``-" otherwise). The exponential can be implemented with the following circuit \cite{nielsen_chuang}

\begin{equation}
 \Qcircuit @C=0.7em @R=0.1em @!R {
   & \ctrl{1} & \qw & \qw & \qw & \qw & \qw & \qw & \qw & \ctrl{1} & \qw \\
   & \targ & \ctrl{1} & \qw & \qw & \qw & \qw & \qw & \ctrl{1} & \targ & \qw \\
   & \qw & \targ & \qw & \qw & \qw & \qw & \qw & \targ & \qw & \qw \\
   & & & \vdots & & & & \vdots \\
   & \qw & \qw & \qw & \ctrl{1} & \qw & \ctrl{1} & \qw & \qw & \qw & \qw \\
   & \qw & \qw & \qw & \targ & \gate{R_{z}(-2\tau)} & \targ & \qw & \qw & \qw & \qw                                
 }
\end{equation}

\noindent
where for the $z$-rotations holds 

\begin{equation}
  R_z(\theta) = \left(\begin{array}{cc} e^{-i\theta/2} & 0 \\ 0 & e^{i\theta/2} \end{array}\right)
\end{equation}

\noindent
and CNOTs assure the correct sign of the phase shift according to the parity of the state.

Due to the following change-of-basis identities \cite{nielsen_chuang}

\begin{eqnarray}
  \sigma_x & = & H \sigma_z H^{\dagger} \\
  \sigma_y & = & Y \sigma_z Y^{\dagger}, 
\end{eqnarray}

\noindent
where

\begin{equation}
  Y =  R_x(-\pi/2) = \frac{1}{\sqrt{2}} \left(\begin{array}{cc} 1 & i \\ i & 1 \end{array}\right),
\end{equation}

\noindent
the exponentials 

\begin{equation}
  \mathrm{exp} \Big[ \frac{i h_{pq}^{\rm{R}} \tau}{2N} \sigma_{x}^{p} \otimes \big( \sigma_{z}^{p \rightarrow q} \big) \otimes \sigma_{x}^{q} \Big] \nonumber 
\end{equation}
\vskip -0.2cm
\begin{equation}  
  \mathrm{exp} \Big[ \frac{i h_{pq}^{\rm{R}} \tau}{2N} \sigma_{y}^{p} \otimes \big( \sigma_{z}^{p \rightarrow q} \big) \otimes \sigma_{y}^{q} \Big] \nonumber 
\end{equation}
\vskip -0.2cm
\begin{equation}  
  \mathrm{exp} \Big[ \frac{i h_{pq}^{\rm{I}} \tau}{2N} \sigma_{y}^{p} \otimes \big( \sigma_{z}^{p \rightarrow q} \big) \otimes \sigma_{x}^{q} \Big] \nonumber 
\end{equation}
\vskip -0.2cm
\begin{equation}  
  \mathrm{exp} \Big[ \frac{-i h_{pq}^{\rm{I}} \tau}{2N} \sigma_{x}^{p} \otimes \big( \sigma_{z}^{p \rightarrow q} \big) \otimes \sigma_{y}^{q} \Big] 
  \label{exponentials}
\end{equation}
\vskip 0.2cm

\noindent
can be implemented with the following circuit pattern,

\begin{equation}
 \Qcircuit @C=0.7em @R=0.1em @!R {
   \lstick{p} & \gate{A^{\dagger}} & \ctrl{1} & \qw & \qw & \qw & \qw & \qw & \ctrl{1}  & \gate {A} & \qw \\
   \lstick{p-1} & \qw & \targ & \qw & \qw & \qw & \qw & \qw & \targ & \qw & \qw \\
   & & & \vdots & & & & \vdots \\ 
   \lstick{q+1}& \qw & \qw & \qw & \ctrl{1} & \qw & \ctrl{1} & \qw & \qw & \qw & \qw \\
   \lstick{q} & \gate{B^{\dagger}} & \qw & \qw & \targ & \gate{R_{z}(\theta)} & \targ & \qw & \qw & \gate{B} & \qw                                 
 }
\label{h1_circuit}
\end{equation}

\vskip 0.3cm
\noindent
where $A$ and $B$ are for the individual exponentials (\ref{exponentials}) equal to $\{ H,H\}$, $\{ Y,Y\}$, $\{ Y,H\}$, and $\{ H,Y\}$, respectively, and $\theta$ to $-h_{pq}^{\mathrm{R}}\tau/N$, $-h_{pq}^{\mathrm{R}}\tau/N$, $-h_{pq}^{\mathrm{I}}\tau/N$, and $h_{pq}^{\mathrm{I}}\tau/N$, respectively. Note that although the two strings of $\sigma$ matrices in the first parenthesis in (\ref{off_diag}) commute as do the two strings in the second parenthesis, they do not commute mutually. This, however, is not a complication since the Trotter approximation (\ref{trotter}) must be employed anyway.

We have demonstrated the decomposition technique for the \textit{direct mapping} approach on the one-electron part of the Hamiltonian. The procedure for the two-electron part is more elaborate, but completely analogous and we refer an interested reader to \cite{whitfield_2010}, where all the cases are presented in a systematic way.

The overall scaling of the algorithm is given by the scaling of a single controlled action of the unitary propagator without repetitions enforced by the Trotter approximation (\ref{trotter}). These repetitions increase only the prefactor to the polynomial scaling, not the scaling itself. Also the required precision is limited, about 20 binary digits of $\phi$ are sufficient to achieve the chemical accuracy \cite{veis_2010}. 

\begin{figure}[!ht]
 \begin{center}
   \includegraphics[width=12cm]{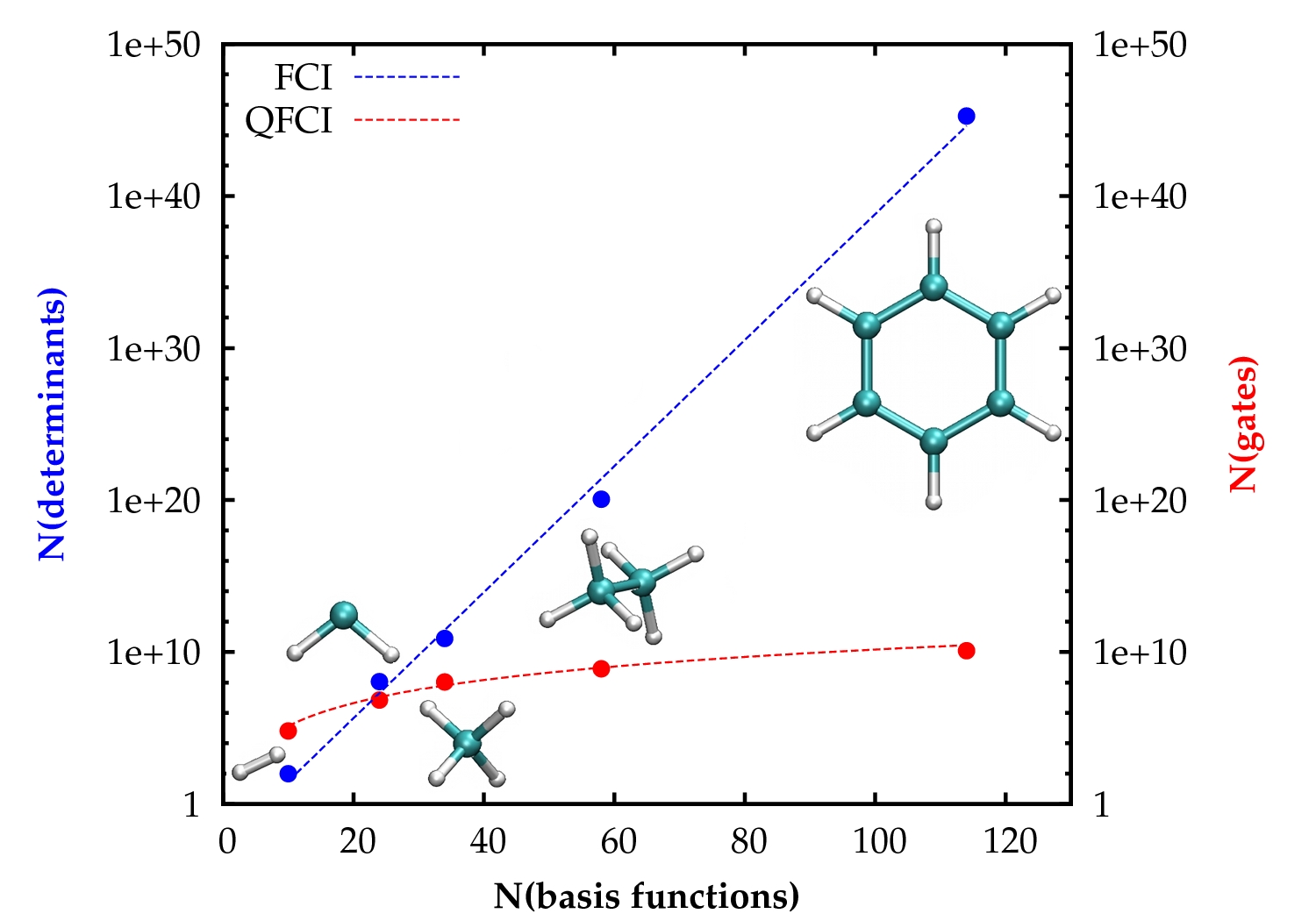}
   \caption{The \textit{exponential} speedup of the qFCI over the FCI. In case of the FCI (blue), dependence of the number of Slater determinants in the FCI expansion on the number of basis functions is shown. In case of the qFCI (red), dependence of the number of one and two qubit gates needed for a single controlled action of the unitary operator on the number of basis functions is presented. Points in the graph correspond to the depicted molecules (hydrogen, methylene, methane, ethane, and benzene) in the cc-pVDZ basis set. Reprinted from \cite{veis_2010}.} 
   \label{scaling}    
 \end{center} 
\end{figure}

The single controlled action of the exponential of a one-body Hamiltonian (\ref{h1}) results in $\mathcal{O}(n^3)$ scaling: there are $\mathcal{O}(n^2)$ different $h_{pq}$ terms and each of them requires $\mathcal{O}(n)$ elementary quantum gates [see the circuit (\ref{h1_circuit})]. Since the same decomposition technique applied to the two-electron part of the Hamiltonian gives rise to similar circuit patterns \cite{whitfield_2010}, each term $g_{pqrs}$ requires $\mathcal{O}(n)$ elementary quantum gates as well (this in fact holds for general $m$-body Hamiltonians \cite{ovrum_2007}). The total scaling is thus $\mathcal{O}(n^5)$ \cite{lanyon_2010, whitfield_2010}, where $n$ is the number of molecular spin orbitals (or bispinors in the relativistic case \cite{veis_2011}) and the qFCI achieves an \textit{exponential} speedup over the conventional FCI. This speedup is demonstrated in Figure \ref{scaling}.

At this point, we would like to make few remarks. Firstly, we assumed that the initial state preparation is an \textit{efficient} step, as was already mentioned. Secondly, when a quantum chemical method with a scaling worse than $\mathcal{O}(n^5)$ is used for calculation of the initial guess state on a conventional computer, then this classical step becomes a rate determining one. Besides this, the classical computation of the integrals in the molecular orbital basis scales as $\mathcal{O}(n^{5})$ (due to the integral transformation) as well. We also assumed noise-free qubits and thus did not take into account any quantum error correction \cite{gaitan_book}. Clark et al. studied the resource requirements for a similar, but fault-tolerant computation of the ground state of a one dimensional transverse Ising model \cite{clark_2009} on a proposed scalable quantum computing architecture \cite{metodi_2005}. They showed that due to the exponential scaling of the resource requirements with the desired energy precision as well as due to the Trotter approximation employed, an elaborate error correction is required, which leads to a huge increase of a computational time. They also gave the values of the experimental parameters (e.g. the physical gate time) needed for acceptable computational times. However, the question of reducing the resource requirements needed for fault tolerant qFCI computations is still open and undergoes an \mbox{active research}.

\section{Application to non-relativistic molecular Hamiltonians}
\label{NR}

We will demonstrate an application of the qFCI method to non-relativistic ground and excited state energy calculations on the example of the methylene molecule \cite{veis_2010}.

\subsection{Example of CH$_{2}$ molecule}
Methylene molecule (CH$_{2}$) in a minimal basis set (STO-3G) is a simple, yet computationally interesting system suitable for simulations and performance testing of the qFCI method. CH$_{2}$ is well known for the multireference character of its lowest-lying singlet electronic state ($\tilde{a}~^{1}A_{1}$) and is often used as a benchmark system for testing of newly developed computational methods (see e.g. \cite{bwcc1, evangelista-allen1, mkcc_our2, demel-pittner-bwccsdt}). In the STO-3G basis set, the total number of molecular (spin)orbitals is 7(14) which means 15 qubits in the direct mapping approach (one qubit is needed in the read-out part of the register).
Since classical simulations of 15-qubit computations are feasible, the CH$_2$ molecule serves as an excellent candidate for the first benchmark simulations. 

\begin{figure}[!ht]
  \begin{center}
    \subfloat[][]{  
      \hskip -1cm
      \includegraphics[width=8.5cm]{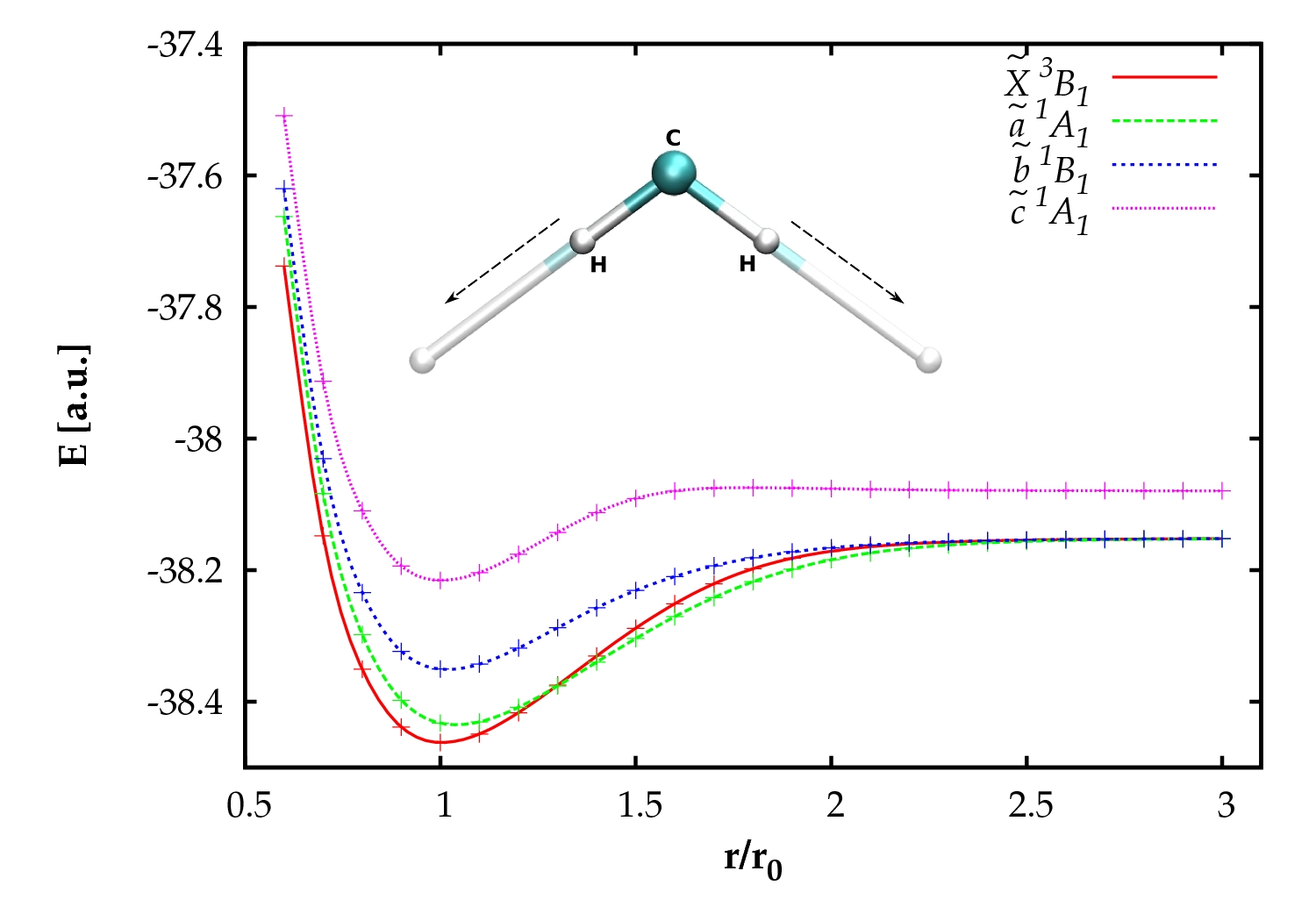}
      \label{stretch}
    } 
    \subfloat[][]{
      \includegraphics[width=8.5cm]{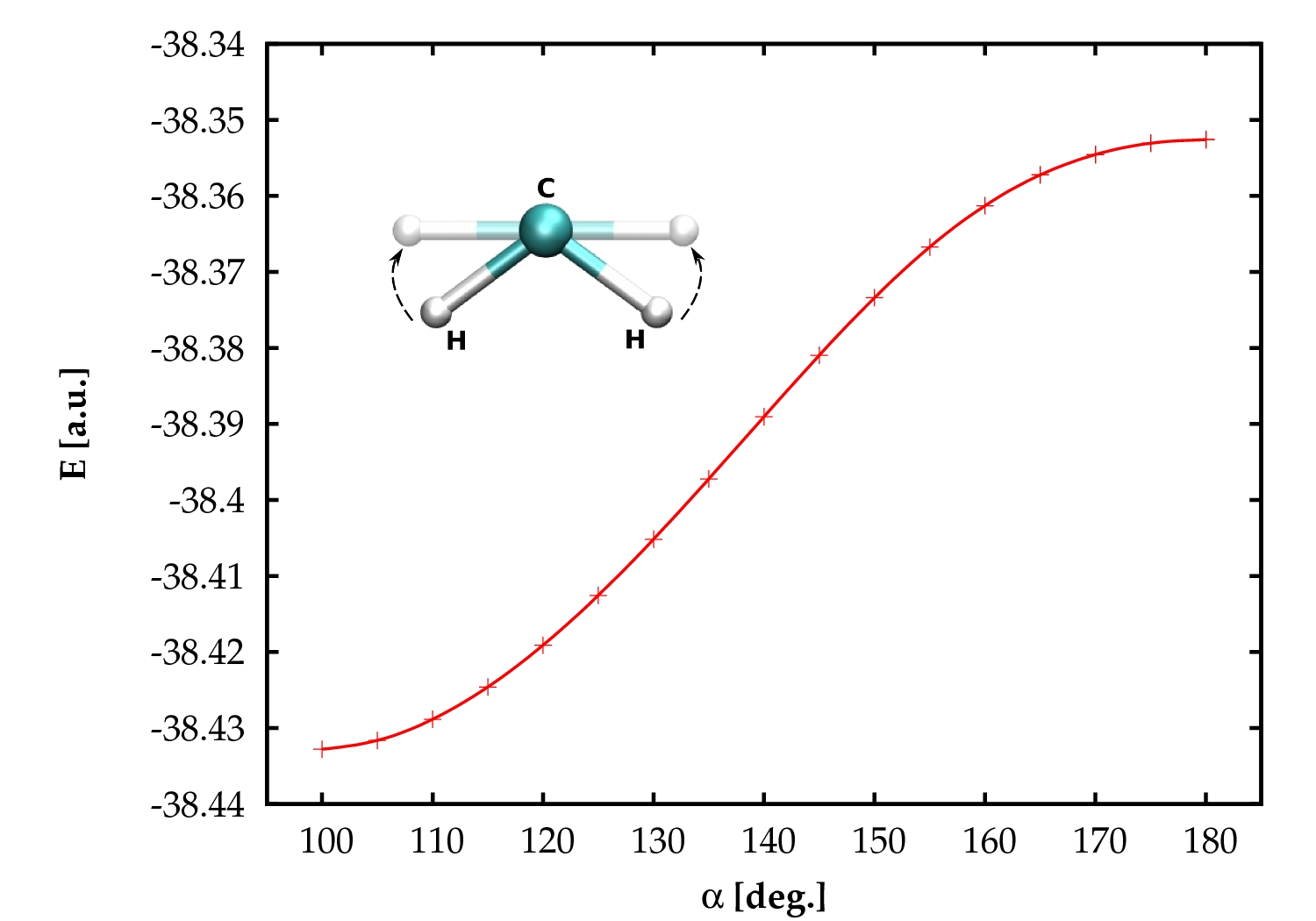}
      \label{bend}
    }   
    \caption{(a) Energies of the four simulated states of CH$_{2}$ for the C-H bond stretching, $r_{0}$ denotes the equilibrium bond distance. (b) Energy of $\tilde{a}~^{1}A_{1}$ state of CH$_{2}$ for the H-C-H angle bending, $\alpha$ denotes the H-C-H angle. Figures reprinted from \cite{veis_2010}.}
    \label{ch2}
  \end{center}
\end{figure}

We simulated the qFCI energy calculations of the four lowest-lying electronic states of CH$_{2}$: $\tilde{X}~^{3}B_{1}$, $\tilde{a}~^{1}A_{1}$, $\tilde{b}~^{1}B_{1}$, and $\tilde{c}~^{1}A_{1}$. Two processes shown in Figure \ref{ch2} were studied: C-H bond stretching (both C-H bonds were stretched, Figure \ref{stretch}), and H-C-H angle bending for $\tilde{a}~^{1}A_{1}$ state (Figure \ref{bend}). These processes were chosen designedly because description of bond breaking is a hard task for many of computational methods and H-C-H angle bending since the $\tilde{a}~^{1}A_{1}$ state exhibits a very strong multireference character at linear geometries. 

Our work followed up the work by Wang et al. \cite{wang_2008} who studied the influence of initial guesses on the performance of the quantum FCI method on two singlet states of water molecule across the bond-dissociation regime. They found out that the Hartree-Fock (HF) initial guess is not sufficient for bond dissociation and suggested the use of CASSCF method. Few configuration state functions added to the initial guess in fact improved the success probability dramatically.

We therefore also tested different initial guesses for the qFCI calculations. Those denoted as HF guess were composed only from spin-adapted configurations which qualitatively describe certain state. Here, for simplicity, we will present the results just for the $\tilde{X}~^{3}B_{1}$ ground state described by the configuration $(1a_{1})^{2}(2a_{1})^{2}(1b_{2})^{2}(3a_{1})(1b_{1})$ and $\tilde{a}~^{1}A_{1}$ state characterized by $(1a_{1})^{2}(2a_{1})^{2}(1b_{2})^{2}(3a_{1})^{2}$ configuration. Initial guesses denoted as CAS($x$,$y$) were based on complete active space configuration interaction (CASCI) calculations with small active spaces, which contained $x$ electrons in $y$ orbitals. Initial guesses were constructed only from the configurations whose absolute values of amplitudes were higher than 0.1. Those constructed from the configurations whose absolute values of amplitudes were higher than 0.2 are denoted as CAS($x$,$y$), tresh. 0.2 guess. All the initial guesses were normalized before the simulations.

In our simulations, similarly as in \cite{aspuru-guzik_2005}, the exponential of a Hamiltonian operator was implemented as an $n$-qubit gate. This is motivated by the fact that once the decoherence is not involved in the model, the exponential of a Hamiltonian can be obtained with an arbitrary precision simply by a proper number of repetitions in (\ref{trotter}). But we would like to note that Whitfield et al. \cite{whitfield_2010} among others also numerically studied the length of a Trotter time step needed for a required energy precision on the example of the Helium atom. We ran 20 iterations of the IPEA which correspond to the final energy precision $\approx 10^{-6}~E_h$. All the other computational details including the exact definition of the CA spaces can be found in \cite{veis_2010}.

\begin{figure}[!h]
 \begin{center}
   \includegraphics[width=10cm]{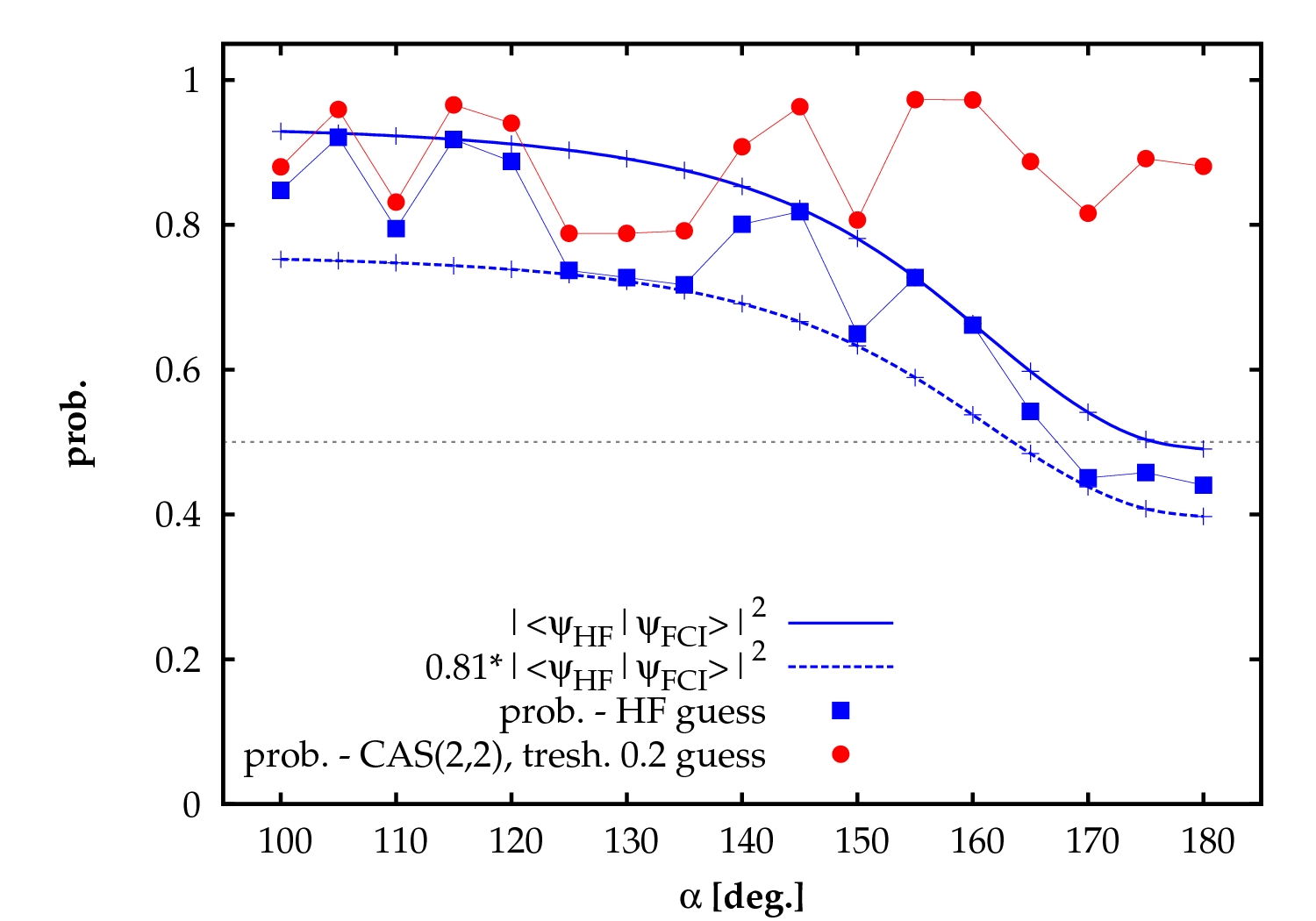}
   \caption{Success probabilities of the \textbf{A} version of IPEA for the $\tilde{a}~^{1}A_{1}$ state with HF and CAS(2,2) initial guesses, tresh 0.2 means that only configurations with absolute values of amplitudes higher than 0.2 were involved in the initial guess, $\alpha$ denotes the H-C-H angle. Reprinted from \cite{veis_2010}.} 
   \label{bend_alg0}    
 \end{center} 
\end{figure}

\begin{figure}[!ht]
  \begin{center}
    \subfloat[][]{  
      \hskip -1cm
      \includegraphics[width=8.5cm]{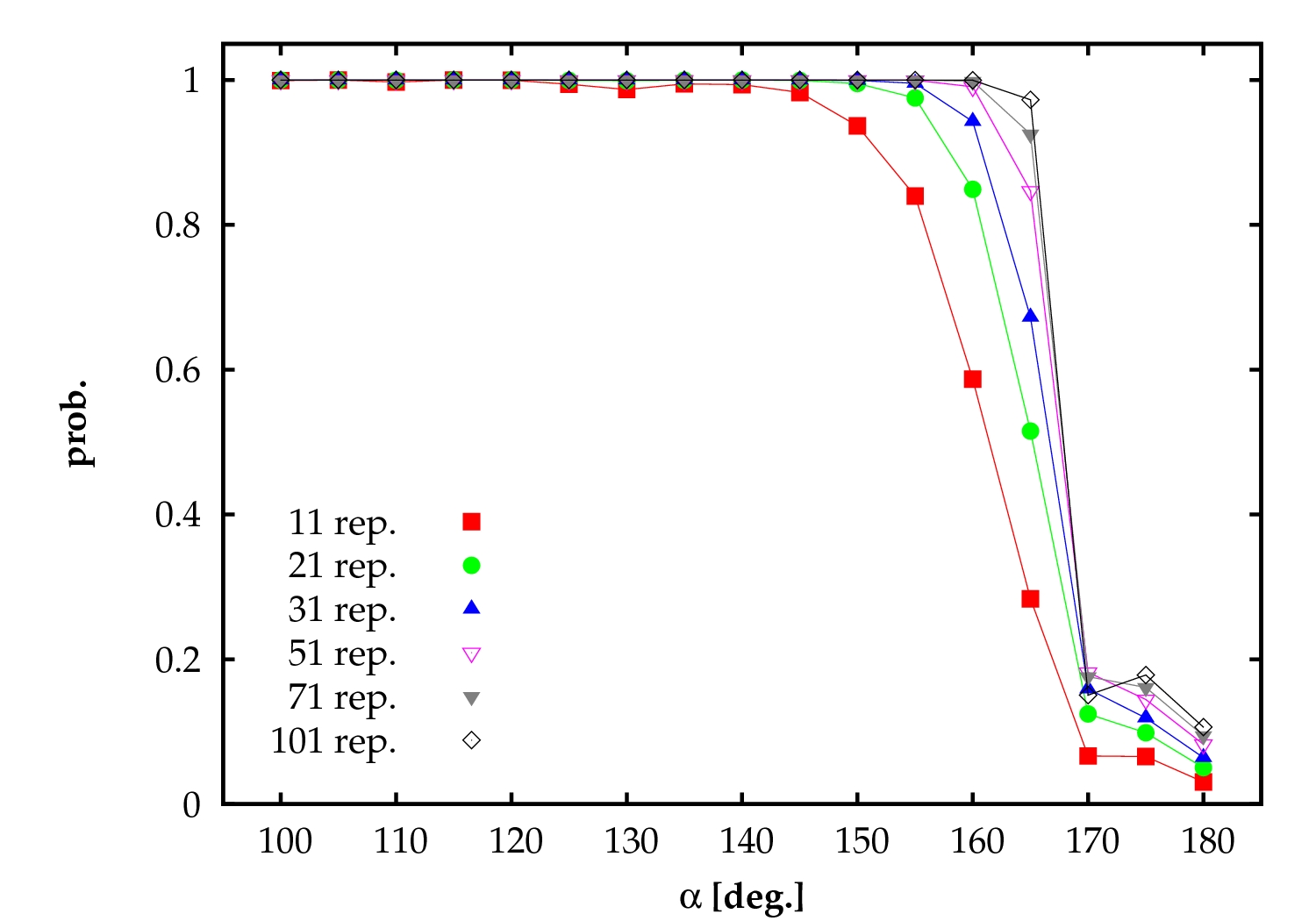}
      \label{bend_rep_hf}
    } 
    \subfloat[][]{
      \includegraphics[width=8.5cm]{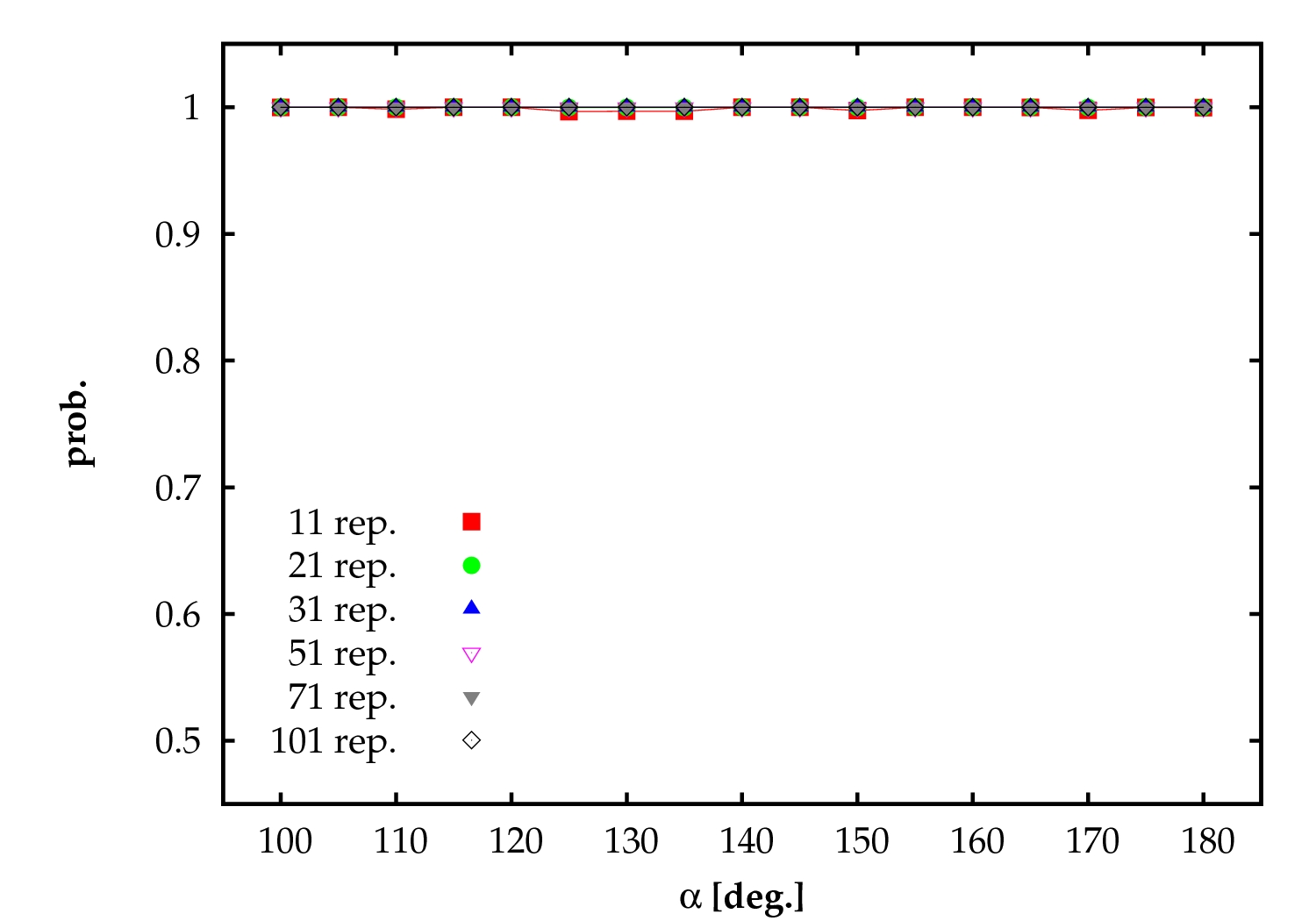}
      \label{bend_rep_cas}
    }   
    \caption{Success probabilities of the \textbf{B} version of IPEA with (a) HF and (b) CAS(2,2), tresh. 0.2 initial guesses and different number of repetitions of individual bit measurements for the $\tilde{a}~^{1}A_{1}$ state, $\alpha$ denotes the H-C-H angle. Figures reprinted from \cite{veis_2010}.}
    \label{bend_rep}
  \end{center}
\end{figure}

Figure \ref{bend_alg0} presents the results of the \textbf{A} version of IPEA for the angle bending process. The overlap and scaled overlap of the initial HF guess wave function and the exact FCI wave function is shown as well as the success probabilities for the HF and CAS(2,2) initial guesses and dotted line bounding the safe region. The results numerically confirm that the success probabilities always lie in the interval $\big| \langle \psi_{\rm{init}} | \psi_{\rm{exact}} \rangle \big|^{2} \cdot \big( 0.81, 1 \big>$, depending on the value of the remainder $\delta$ (\ref{reminder}). This algorithm can be \textit{safely} used when the resulting success probability is higher than 0.5 (as it can then be amplified by repeating the whole process). When going to the linear geometry, where the $\tilde{a}~^{1}A_{1}$ state exhibits a very strong multireference character and the restricted HF (RHF) description is no more adequate, the CAS(2,2) initial guess improves the success probabilities dramatically. Moreover, these initial states correspond to only two configurations and are thus very easy to prepare \cite{ortiz_2001}.

Performance of the \textbf{B} version of IPEA with the HF and CAS(2,2) initial guesses is illustrated in Figure \ref{bend_rep}. As can be seen, in the situations where the particular initial guess state can be used, few repetitions are enough to amplify the success probability to unity. The HF initial guess is again not sufficient for the linear and close to linear geometries.

\begin{figure}[!ht]
  \begin{center}
    \subfloat[][]{  
      \hskip -1cm
      \includegraphics[width=8.5cm]{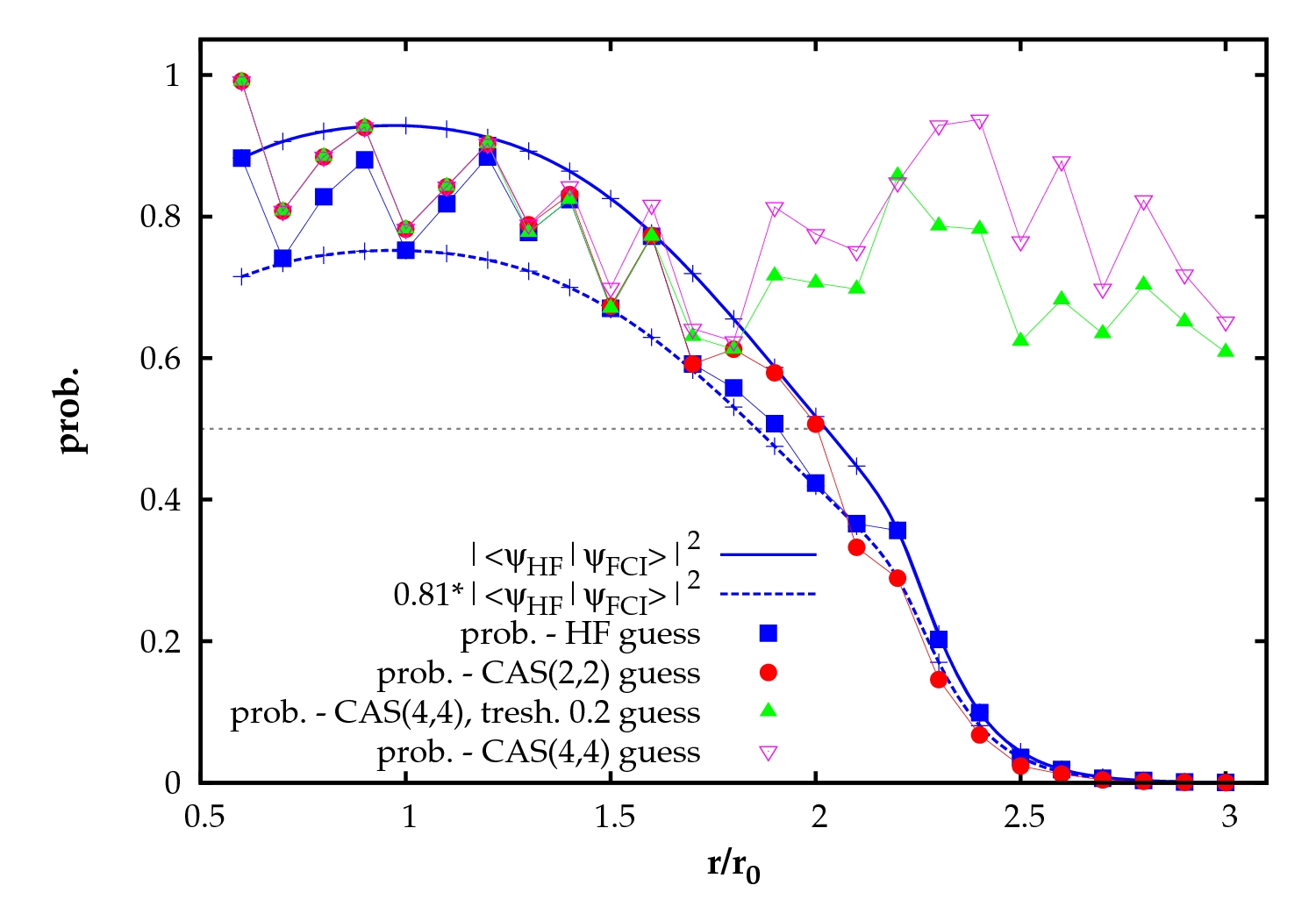}
      \label{stretch_singlet}
    } 
    \subfloat[][]{
      \includegraphics[width=8.5cm]{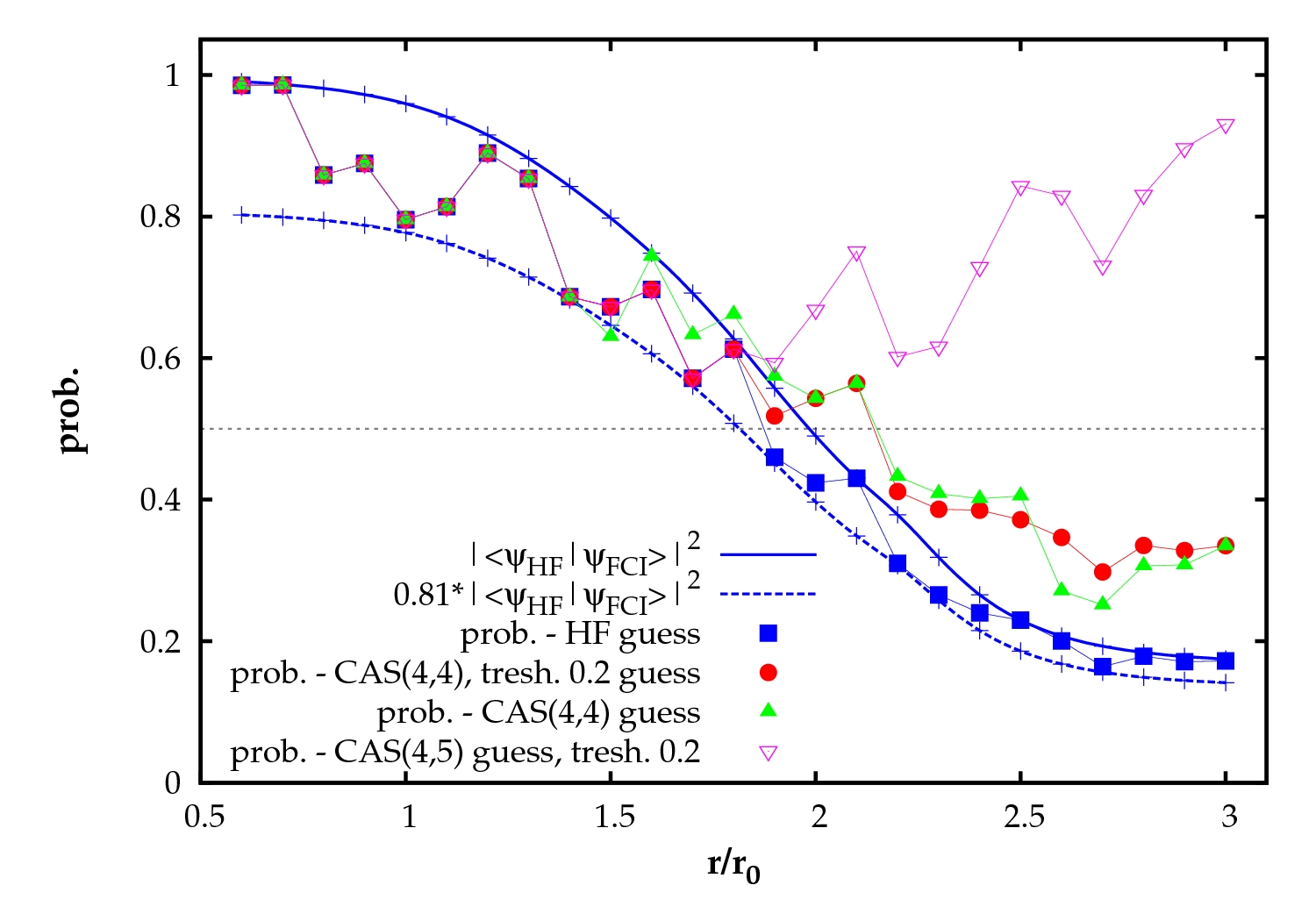}
      \label{stretch_triplet}
    }   
    \caption{Success probabilities of the \textbf{A} version of IPEA for (a) $\tilde{a}~^{1}A_{1}$ state, (b) $\tilde{X}~^{3}B_{1}$ state of CH$_2$. Different initial guesses were used, tresh 0.2 means that only configurations with absolute values of amplitudes higher than 0.2 were involved in the initial guess, $r_{0}$ denotes the equilibrium bond distance. Figures reprinted from \cite{veis_2010}.}
    \label{stretch_alg0}
  \end{center}
\end{figure}


\begin{figure}[!ht]
  \begin{center}
    \subfloat[][]{  
      \hskip -1cm
      \includegraphics[width=8.5cm]{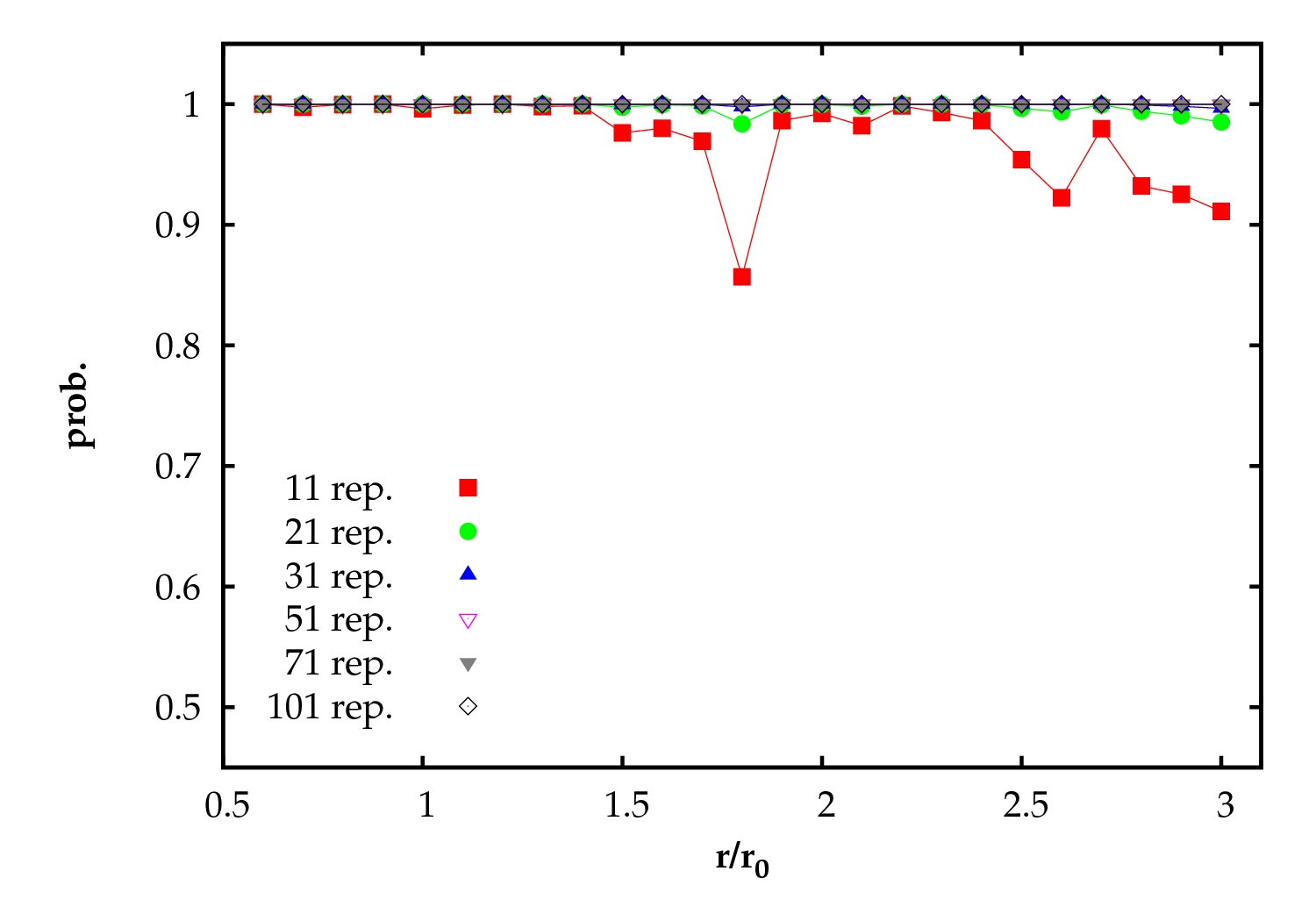}
      \label{stretch_rep_singlet}
    } 
    \subfloat[][]{
      \includegraphics[width=8.5cm]{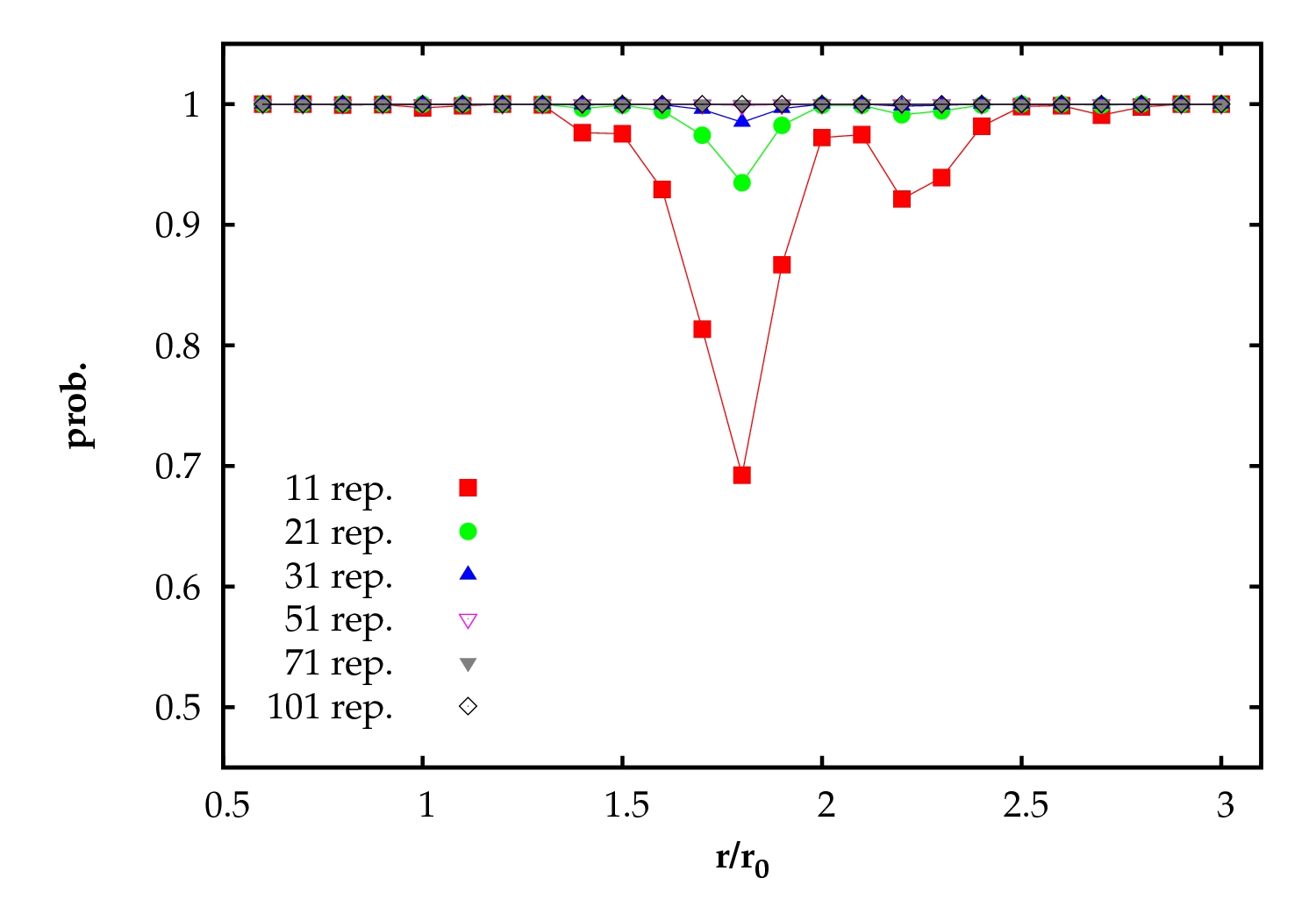}
      \label{stretch_rep_triplet}
    }   
    \caption{Success probabilities of the \textbf{B} version of IPEA with ``best" initial guesses and different number of repetitions of individual bit measurements, $r_{0}$ denotes the equilibrium bond distance. (a) $\tilde{a}~^{1}A_{1}$ state, CAS(4,4), tresh. 0.2 guess, (b) $\tilde{X}~^{3}B_{1}$ state, CAS(4,5), tresh. 0.2 guess. Figures reprinted from \cite{veis_2010}.}
    \label{stretch_rep}
  \end{center}
\end{figure}

Results corresponding to the $\tilde{a}~^{1}A_{1}$ and $\tilde{X}~^{3}B_{1}$ states for the C-H bond stretching are summarized in Figures \ref{stretch_alg0} and \ref{stretch_rep}. Figure \ref{stretch_alg0} presents the performance of the \textbf{A} version of IPEA. When going to more stretched C-H bonds, the RHF initial guess again fails. The CAS(2,2) initial guess improves the success probabilities in case of $\tilde{a}~^{1}A_{1}$ state near the equilibrium geometry but in the region of more stretched C-H bonds it also fails. In this region, CAS(4,4) initial guess must be used [CAS(4,4), tresh. 0.2 guess is sufficient]. \mbox{The situation} is even more difficult for the $\tilde{X}~^{3}B_{1}$ state when the C-H bonds are stretched. Here, even the CAS(4,4) initial guess fails and bigger active space - CAS(4,5) - must be used for initial guess state calculations. However, apart from the CAS size, initial guess states always contained at most 12 configurations, but usually 8 or even less for nearly dissociated molecule. This observation is in agreement with the results of \cite{wang_2008}, where few configuration state functions added to the initial guess improved the success probability dramatically.

Figure \ref{stretch_rep} presents the performance of the \textbf{B} version of IPEA for the ``best" initial guesses in terms of price/performance ratio. The evident result is again that relatively small number of repetitions ($\approx 31$) amplifies the success probability nearly to unity. This is very important, because the \textbf{B} version of IPEA seems to be a better candidate for the first real larger-scale qFCI calculations, since it does not require a long coherence time. 

\section{Extension to relativistic molecular Hamiltonians}
\label{R}

So far, we concerned non-relativistic computations only. But it is well known that relativistic effects can be very important in chemistry. In fact, accurate description of systems with heavy elements requires adequate treatment of relativistic effects \cite{hess_2000}. The most rigorous approach [besides the quantum electrodynamics (QED) which is not feasible for quantum chemical purposes] is the four component (4c) formalism \cite{dyall}. Very recently, we therefore developed the qFCI method for 4c relativistic computations \cite{veis_2011} and the details of this method are the subject of this section.

The 4c formalism brings three major computational difficulties: (1) working with 4c orbitals (bispinors), (2) complex algebra when molecular symmetry is low, and (3) rather large Hamiltonian matrix eigenvalue problems [due to larger mixing of states than in the non-relativistic case]. All of them can nevertheless be solved on a quantum computer. Before discussing how this is done, we would like to note that in our work, we restricted ourselves to the 4c Dirac-Coulomb Hamiltonian (with the rest mass term $mc^2$ subtracted):

\vskip -0.3cm
\begin{align}
  \hat{H} = \sum_{i=1}^N \left [c(\boldsymbol\alpha_i \cdot \mathbf{p}_i) + \beta^{\prime}_i m c^2 + \sum_A \frac{1}{r_{iA}} \right]
+ \sum_{i<j} \frac{1}{r_{ij}} \\
  \alpha_k = \left( \begin{array}{cc}  0 &\sigma_k \\ \sigma_k &0\end{array} \right), \quad \beta^{\prime} = \beta - I_{4}, \quad \beta =  \left( \begin{array}{c c}  I_2 &0 \\ 0& -I_2\end{array} \right), \nonumber
\end{align}

\noindent
where $\sigma_k$ ($k = x,y,z$) are Pauli matrices and $I_2$ the $2 \times 2$ unit matrix. This type of Hamiltonian covers the major part of the spin-orbit interaction (including two-electron spin-own orbit) and also scalar relativistic effects. It is in fact without loss of generality sufficient for our purposes since going to Dirac-Coulomb-Breit Hamiltonian \cite{dyall} correct to $\mathcal{O}(c^{-2})$ is conceptually straightforward as the inclusion of the corresponding integrals requires a classically polynomial effort.

We will start the description of the algorithm with a mapping of a \textit{relativistic} quantum chemical wave function onto a quantum register. We have already briefly mentioned this topic in Section \ref{Quantum full configuration interaction method}. We in fact do not need any qubits to represent positronic bispinors in the direct mapping due to the no-pair approximation \cite{dyall}, which is widely used in relativistic quantum chemistry. In this approach, one actually builds up an $N$-electron wave function only from Slater determinants containing positive-energy bispinors. This procedure neglects all QED effects, but it is justifiable at the energy scale relevant to chemistry. Moreover, because of the time-reversal symmetry of the Dirac equation, bispinors occur in degenerate Kramers pairs \cite{dyall} denoted $A$ and $B$ (analogy to $\alpha$ and $\beta$ spin in non-relativistic treatment). The direct mapping thus employs one qubit for bispinor $A$ and one for $B$. The 4c character of molecular bispinors does not complicate the approach substantially, since the Hartree-Fock (HF) part of a calculation is done on a classical computer and only the exponentially scaling FCI part on a quantum one.

Assuming the no-pair approximation and the empty Dirac picture, the relativistic Hamiltonian has the same second quantized structure (\ref{ham_sec_quant}) as the non-relativistic one. The only difference is that the molecular integrals $h_{pq}$ and $g_{pqrs}$ are now in general complex (and have thus lower index permutation symmetry). This is actually no difficulty for a quantum computer, since our working environment is a complex vector space of qubits anyway and we do the exponential of a complex matrix even if the Hamiltonian is real. Moreover, on the example of the one-electron part of the Hamiltonian [see (\ref{off_diag}) and the circuit (\ref{h1_circuit})], one can see that the decomposition of the unitary propagator $e^{i\tau\hat{H}}$ with complex molecular integrals requires twice as many gates compared to real-valued Hamiltonians \cite{whitfield_2010}, while complex arithmetic on a classical computer requires four times more operations.

The last of the aforementioned difficulties of the 4c formalism is the size of a Hamiltonian matrix eigenvalue problem. When we put the double group symmetry aside and employ Kramers restricted (KR) approach, the relativistic Hamiltonian, unlike the non-relativistic one, mixes determinants with different values of the pseudo-quantum number $M_{K}$

\begin{equation}
  M_{K} = 1/2(N_{A} - N_{B}),
\end{equation}

\noindent
where $N_A$ and $N_B$ denote the number of electrons in $A$ and $B$ Kramers pair bispinors (in the non-relativistic case, $M_K$ corresponds to $M_S$ and Hamiltonian is block diagonal in $M_S$). Therefore, in the case of a closed shell system with $n$ electrons in $m$ molecular orbitals (bispinors), the size of the relativistic Hamiltonian matrix (number of Slater determinants in the FCI expansion) reads


\begin{equation}
  N_{\rm{rel.}} = \sum_{x=0}^{n} \left( \begin{array}{c}  m \\ x \end{array} \right) \left( \begin{array}{c}  m \\ n - x \end{array} \right) = \left( \begin{array}{c}  2m \\ n \end{array} \right).
\end{equation}

\noindent
The last equality is due to the Vandermonde's convolution \cite{knuth_math}. When compared to the non-relativistic case (\ref{nrdets}) and using the Stirling's approximation in the form

\begin{equation}
  \mathrm{ln}~m! \approx \frac{1}{2} \mathrm{ln}~(2\pi m) + m\mathrm{ln}~m - m \qquad \mathrm{for}~m\rightarrow \infty,
\end{equation}

\noindent
the ratio between the relativistic and non-relativistic Hamiltonian matrix sizes is given by the expression

\begin{equation}
  k_{\rm{rel.}/\rm{non-rel.}} = \frac{N_{\rm{rel.}}}{N_{\rm{non-rel.}}} = \Bigg( \frac{\sqrt{\pi (2k - 1)}}{2k}\Bigg) \cdot m^{1/2},
  \label{ratio}
\end{equation}

\noindent
where we set $m = k \cdot n$. 

On a quantum computer on the other hand, this problem does not occur. When employing the \textit{direct mapping} which maps the whole Fock space of the system on the Hilbert space of qubits, the Hamiltonian (\ref{ham_sec_quant}) in fact implicitly works with all possible values of $M_{K}$. The scaling of the relativistic qFCI method is therefore the same as the non-relativistic one, namely $\mathcal{O}(m^{5})$, where $m$ is the number of molecular orbitals/bispinors (see Section \ref{fact}). 

It is quite remarkable and deserves an emphasis that the relativistic qFCI method not only achieves an \textit{exponential} speedup over its classical counterpart, but, as we have just discussed, also has the same cost (in terms of scaling) as its non-relativistic analogue. 

\subsection{Example of SbH molecule}
We will demonstrate the performance of the relativistic qFCI method on the example of the SbH molecule. This molecule is of particular interest to us, since its non-relativistic ground state $^{3}\Sigma^{-}$ splits due to spin-orbit effects into $X~0^{+}$ and $A~1$ states. In the approximate $\lambda\omega$-projection, these states are dominated by $\sigma_{1/2}^2\pi_{1/2}^2\pi_{3/2}^0$ and $\sigma_{1/2}^2\pi_{1/2}^1\pi_{3/2}^1$ configurations. The splitting is truly of ``molecular nature" as it disappears for dissociated atoms and its experimental value is $\Delta E_{\rm{SO}} = 654.97$ cm$^{-1}$ \cite{balasubramanian_1989}. 

\begin{figure}[!ht]
  \centering
  \includegraphics[width=10cm]{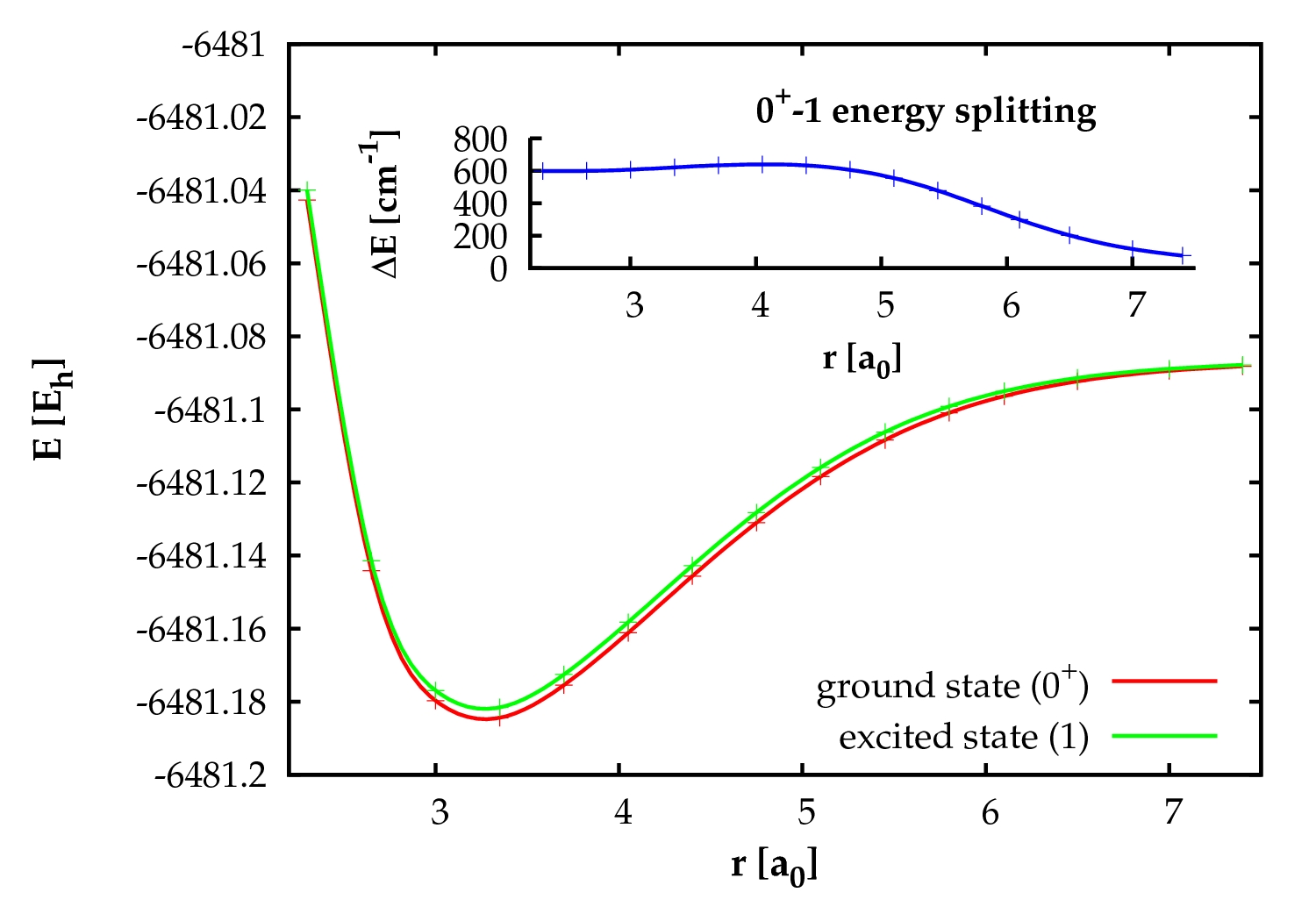}
  \caption{Simulated potential energy curves of ground (0$^{+}$) and excited (1) states of SbH, and spin-orbit energy splitting. Reprinted from \cite{veis_2011}.}
  \label{graf_en}
\end{figure}

For this reason, we simulated the SbH bond dissociation process. Simulated potential energy curves of both states are shown in Figure \ref{graf_en}. Since we employed rather large basis sets of triple-$\zeta$ quality, we could not manage to simulate the FCI calculations with all electrons. We instead simulated general active space (GAS) KRCI computations \cite{fleig_2003} with the occupation constraints that give rise to CI spaces of approximately 30000 determinants. Definition of the GA spaces together with all the computational details can be found in \cite{veis_2011}.

We worked solely with a compact mapping employing the double-group symmetry ($C_{2v}^{*}$) and exponential of a Hamiltonian was again simulated as an $n$-qubit gate. We used the DIRAC program \cite{dirac} for calculations of Hamiltonian matrices and ran 17 iterations of the IPEA which correspond to the final energy precision $\approx$$3.81 \times 10^{-6}$ $E_h$.

\begin{figure}[!ht]
  \begin{center}
    \subfloat[][]{  
      \hskip -1cm
      \includegraphics[width=8.5cm]{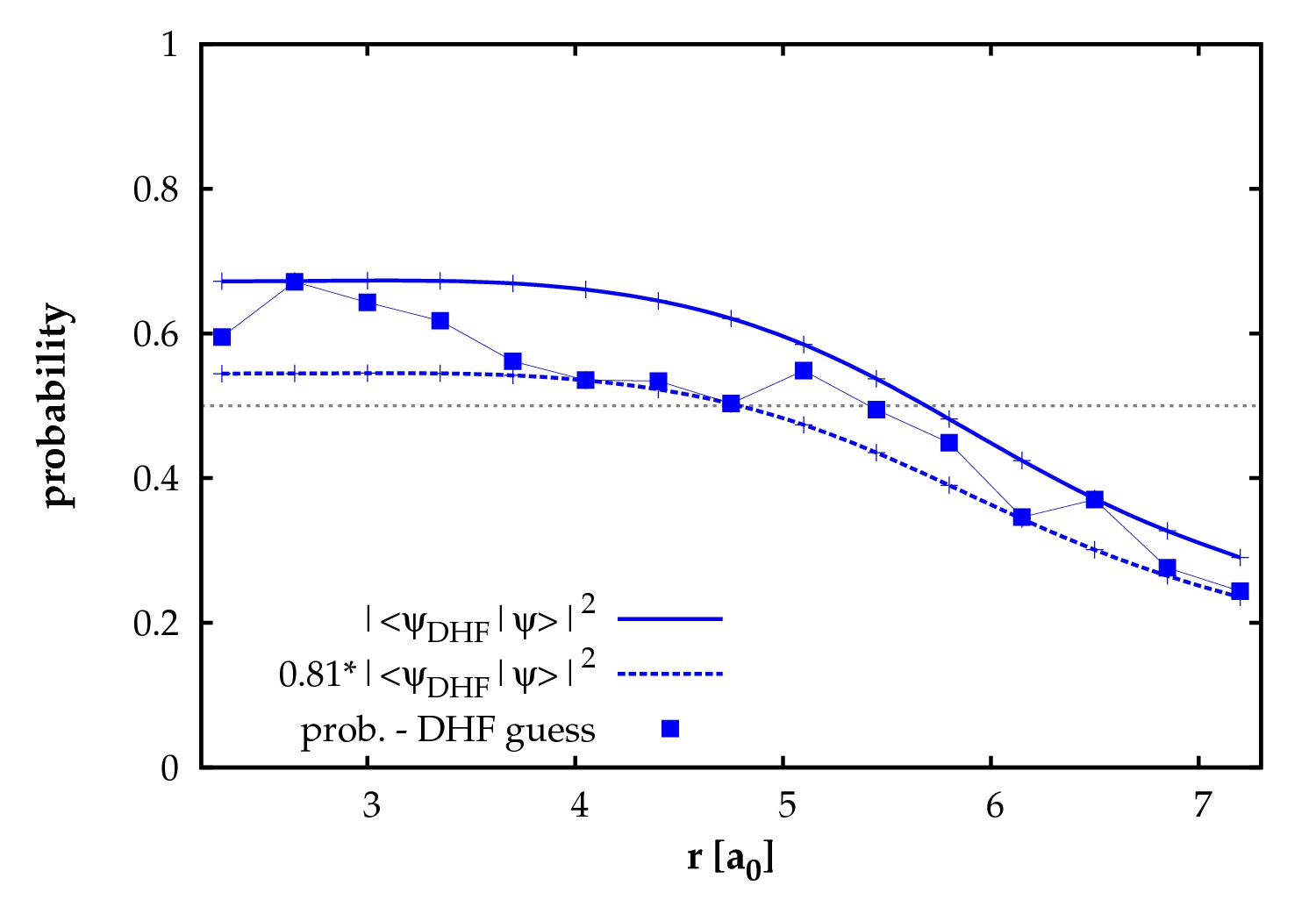}
    } 
    \subfloat[][]{
      \includegraphics[width=8.5cm]{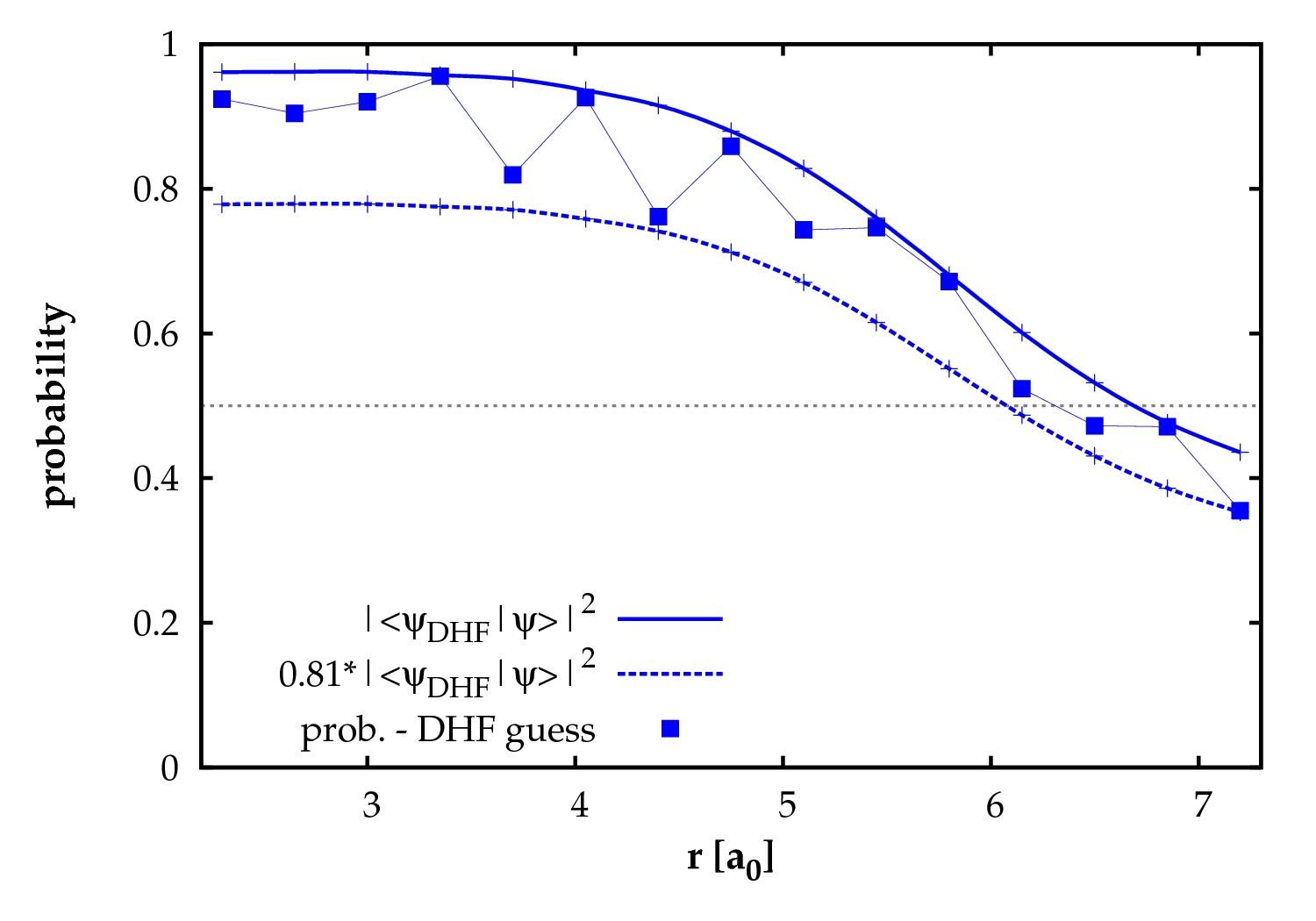}
    }   
    \caption{SbH (a) ground (0$^{+}$) and (b) excited (1) state qFCI (IPEA version \textbf{A}) success probabilities corresponding to the HF initial guesses.}
    \label{prob}
  \end{center}
\end{figure}

Based on our KRCI setup we obtain a vertical $\Delta E_{\rm{SO}}$ of 617 cm$^{-1}$. Success probabilities of the algorithm with the HF initial guesses (represented by the dominant configurations of both states) are presented in Figure \ref{prob}. They correspond to the \textbf{A} version of IPEA. 


Ground state success probabilities confirm that relativistic states have, due to near degeneracies caused by the spin-orbit coupling, often a stronger multireference character than non-relativistic ones. The upper bound of the success probability is less than 0.7 even for the equilibrium geometry and the lower bound is higher than 0.5 only up to 4.8 $a_0$. The success probabilities of the $A~1$ state are higher and the HF initial guesses can be in a noise-free environment used up to 6 $a_0$. For longer distances, initial guesses from more accurate \textit{polynomially} scaling relativistic methods must be used (analogously to the non-relativistic example, e.g. the relativstic 4c CASSCF method with a small orbital CAS). The low ground state success probabilities can be also improved by the ASP procedure as is shown in \mbox{Figure \ref{asp}}.


\section{Conclusions}
In this article, we have reviewed quantum algorithms for the FCI energy calculations. Starting with the seminal paper by Aspuru-Guzik et al. \cite{aspuru-guzik_2005} and ending with our most recent work on relativistic generalization of the algorithm\cite{veis_2011}, we gave a detailed description of the qFCI method which is applicable to the ground as well as excited state energy calculations in the non-relativistic and also relativistic regimes. Both variants share the scaling in the number of molecular orbitals/bispinors $m$, namely $\mathcal{O}(m^5)$ and exhibit thus an \textit{exponential} speedup over their classical counterparts. We have demonstrated their performance by numerical simulations of the CH$_2$ and SbH energy calculations. The results indicate that the ground and excited state energies at equilibrium geometries are accessible with HF initial guesses, which are easy to prepare. On the example of the CH$_2$ molecule, we have shown that CASCI initial guess states with small complete active spaces composed of relatively few configurations ($\approx 10$) are sufficient even for a nearly dissociated molecule to achieve the probability amplification regime of the IPEA algorithm. The first proof-of-principle experimental realizations recently achieved by several groups \cite{lanyon_2010, du_2010, li_2011} are very promising and confirm the real applicability of the method.

\section*{Acknowledgments}
The authors thank Prof. Timo Fleig for fruitful discussions about the relativistic quantum chemistry and gratefully acknowledge the financial support from the Grant Agency of the Czech Republic-GA\v{C}R (203/08/0626) and the Grant Agency of the Charles University-GAUK (114310). L.V. also acknowledges support from the Hl\'{a}vka Foundation.

\bibliography{kvantove_pocitace,clanek_ch2,cc}

\end{document}